\documentclass[10pt,letterpaper]{article}
\usepackage[top=0.85in,left=2.75in,footskip=0.75in]{geometry}

\usepackage{amsmath,amssymb}

\usepackage{changepage}

\usepackage{textcomp,marvosym}

\usepackage{cite}

\usepackage{nameref,hyperref}

\usepackage[right]{lineno}

\usepackage[nopatch=eqnum]{microtype}
\DisableLigatures[f]{encoding = *, family = * }

\usepackage[table]{xcolor}

\usepackage{array}

\newcolumntype{+}{!{\vrule width 2pt}}

\newlength\savedwidth

\newcommand\thickhline{\noalign{\global\savedwidth\arrayrulewidth\global\arrayrulewidth 2pt}%
\hline
\noalign{\global\arrayrulewidth\savedwidth}}

\raggedright
\usepackage[aboveskip=1pt,labelfont=bf,labelsep=period,justification=raggedright,singlelinecheck=off]{caption}

\makeatletter
\renewcommand{\@biblabel}[1]{\quad#1.}
\makeatother

\usepackage{lastpage,fancyhdr,graphicx}
\usepackage{epstopdf}
\usepackage{float}

\fancyheadoffset[L]{2.25in}
\fancyfootoffset[L]{2.25in}
\begin{document}
\vspace*{0.2in}

\begin{flushleft}
{\Large
\textbf\newline{Improving \textit{Wolbachia}-Based Control Programs in Urban Settings: Insights from Spatial Modeling} 
}
\newline
\\
Daniela Florez\textsuperscript{1,2*},
Ricardo Cortez\textsuperscript{1},
James M. Hyman\textsuperscript{1},
Zhuolin Qu\textsuperscript{3}
\\
\bigskip
\textbf{1} Department of Mathematics, Tulane University, New Orleans, LA, USA\\
\textbf{2} Department of Biological Sciences, University of Notre Dame, South Bend, IN, USA
\\
\textbf{3} Department of Mathematics, University of Texas at San Antonio, San Antonio, TX, USA
\\
\bigskip

* Corresponding author\\
Email: dflorezp@nd.edu

\end{flushleft}

\section*{Abstract}
Mosquito-borne diseases, such as dengue fever, remain major global public health challenges, particularly in regions experiencing rapid climate change. Traditional vector control methods often lack efficacy and sustainability, highlighting the need for innovative strategies. One promising approach involves releasing \textit{Aedes aegypti} mosquitoes infected with \textit{Wolbachia}, a symbiotic bacterium that reduces mosquito-borne virus transmission. However, spatial heterogeneity hinders large-scale \textit{Wolbachia} deployment, especially in complex urban settings. High spatial variation and restricted access to certain regions often cause establishment failures, wasting resources and disproportionately affecting disadvantaged communities. In light of this, we developed and analyzed a partial differential equation model to simulate the spatial dynamics of \textit{Wolbachia} transmission among mosquito population, incorporating practical pre-release interventions such as insecticide spraying and repeated releases while contextualizing them within diverse urban landscapes influencing mosquito movement. We identified strategies for optimizing \textit{Wolbachia} releases under constraints on release size and the efficacy level of insecticide used for pre-release interventions. Under a release size restriction, in areas with limited mosquito dispersal (e.g., backyards or gated communities), where  \textit{Wolbachia} mosquitoes remain concentrated locally after release, insecticide use may not be essential for successful \textit{Wolbachia} establishment.
In contrast, in regions with higher dispersal rates (e.g., parks or city blocks), pre-release interventions, specifically reducing at least $35\%$ of the wild mosquito population, enhance \textit{Wolbachia} establishment within nine months. When insecticide efficacy is limited to $35\%$, smaller release sizes than the constrained value achieve $90\%$ \textit{Wolbachia} infection in low-dispersal areas. However, larger releases are required in high-dispersal regions to ensure successful \textit{Wolbachia} establishment. Additionally, our simulation results suggest that distributing \textit{Wolbachia} releases in 2-5 weekly batches may be more effective than a single large release, even without pre-release interventions. These findings highlight the potential of tailoring pre-release interventions and release strategies based on local mosquito dispersal characteristics,  
offering actionable insights for cost-effective and efficient \textit{Wolbachia}-based vector control programs.

\section*{Author summary}
Arboviral diseases remain a major public health concern, particularly in tropical and subtropical regions where mosquito populations thrive. One promising strategy to curb transmission is the release of \textit{Aedes aegypti} mosquitoes infected with \textit{Wolbachia}, a bacterium that reduces their ability to spread viruses. However, past large-scale releases have not always been successful, especially in complex urban settings, where restricted access to certain areas often leads to infection establishment failures and wasted resources. To address this, we developed a spatial model that simulates how \textit{Wolbachia}-infected mosquitoes are established in different urban environments. We also explored strategies to improve their success under constraints on release size and the efficacy level of insecticide used for pre-release interventions. Our findings suggest that targeted releases are most effective in areas with limited mosquito movement without additional insecticide use. In higher-dispersal areas, reducing at least $35\%$ of wild mosquitoes before release significantly improves establishment within nine months. Additionally, distributing releases over 2-5 weekly batches enhances success more than a single large release, even without other interventions. These findings offer practical insights for designing cost-effective and efficient \textit{Wolbachia}-based mosquito control programs, reducing the burden of mosquito-borne diseases on vulnerable communities.

\section{Introduction}\label{Introduction}

Mosquito-borne diseases (MBDs) are the most significant contributors to the human vector-borne disease burden, placing over $80\%$ of the global population at risk each year \cite{franklinos2019effect}. Current research indicates that rapidly changing climatic conditions are likely to exacerbate the impact of MBDs on global morbidity and mortality, highlighting the urgent need for effective resource allocation strategies for mitigation and control \cite{manore2014comparing}. Previous studies have demonstrated that \textit{Wolbachia pipientis} (\textit{Wolbachia}) presents a promising strategy for controlling the spread of MBDs.
Since 2011, the World Mosquito Program has collaborated with governments and communities to deploy \textit{Wolbachia}-infected mosquitoes in eleven countries \cite{turner2023economic}. Several randomized and non-randomized field trials conducted in countries such as Yogyakarta, Indonesia, Vietnam, and Australia \cite{utarini2021efficacy,o2018scaled,indriani2020reduced,pinto2021effectiveness} have demonstrated the successful establishment of \textit{w}Mel in local \textit{Aedes aegypti} populations and the effectiveness of this intervention in controlling dengue and other \textit{Aedes}-borne diseases. However, large-scale trials may not always be feasible, particularly in resource-constrained settings, and success has varied depending on local ecological and operational factors.

For instance, in some locations, \textit{w}Mel did not fully establish, as observed in release programs conducted in Rio de Janeiro, Brazil \cite{dos2022estimating}, and Medellín, Colombia \cite{velez2023large}. A recent analysis of the Medellín trial (2017–2022) \cite{calle2024evaluation} examined the prevalence and distribution of \textit{Wolbachia}-infected mosquitoes two years after releases concluded. The study found a decline in \textit{Wolbachia} prevalence, with only $33.5\%$ of female \textit{Aedes aegypti} mosquito pools testing positive. This prevalence varied significantly across communities, highlighting the heterogeneity of \textit{Wolbachia} distribution in Medellín's urban environment \cite{calle2024evaluation}. 
Similarly, in Rio de Janeiro, Brazil, releases conducted between 2017 and 2019 resulted in only $32\%$ of mosquitoes from release zones testing positive for \textit{w}Mel during the first 29 months. The factors underlying this limited establishment remain unclear despite the large number of releases \cite{dos2022estimating}. These examples emphasize the challenges of achieving consistent success in different urban and ecological contexts, particularly when resources for sustained releases or monitoring are limited.

Drawing on insights from previous trials where \textit{Wolbachia} establishment faced challenges, we develop a mathematical model to guide strategies that increase the efficiency of \textit{Wolbachia} releases in urban settings while addressing practical constraints, such as limited mosquito quantities and the efficacy of pre-release insecticide use. Our model accounts for mosquito dispersal heterogeneity, which is often overlooked in prior models but plays a crucial role in determining \textit{Wolbachia} persistence. To improve establishment success, we evaluate a population replacement approach that requires the fraction of \textit{Wolbachia}-infected mosquitoes to exceed a critical threshold for the infection to persist. Our model specifically assesses pre-release interventions, such as thermal fogging, to reduce the wild mosquito population before \textit{Wolbachia} release, identifying their role in enhancing establishment under different mosquito dispersal scenarios.

Existing mathematical models for predicting \textit{Wolbachia} spread in wild mosquito populations often lack spatial heterogeneity, which limits their applicability to field trials and large-scale mitigation efforts. More precisely, their mechanistic approach of \textit{Wolbachia} transmission ignores the impact of heterogeneous spatial distributions of infected mosquitoes on establishing an endemic state of infection. In these models, the critical threshold for the proportion of \textit{Wolbachia}-infected mosquitoes required to initiate a wave of infection is based on the assumption that both infected and uninfected mosquito populations are homogeneously mixed in space \cite{qu2018modeling}. However, even if the initial release of \textit{Wolbachia}-infected mosquitoes exceeds this threshold near the release site, it may fall short near the edges of the release area. This highlights the need to extend existing ordinary differential equation (ODE) models to partial differential equation (PDE) models that account for spatial variation in mosquito dynamics \cite{qu2022modeling}.

Given the challenges of analyzing complex, high-dimensional partial differential equations (PDEs), many earlier spatial models were developed using heuristics and strong assumptions to produce physically plausible solutions. In \cite{barton2011spatial}, a reaction-diffusion spatial model was introduced, accounting for \textit{Wolbachia}-induced cytoplasmic incompatibility (CI) and fitness cost. The authors used a cubic approximation to model \textit{Wolbachia} vertical transmission and observed traveling wave solutions in this simplified heuristic framework. They also illustrated and derived the threshold introduction size needed to initiate the wave. In \cite{lewis1993waves}, a two-equation spatial model was proposed for a different biological control method, where sterilized insects are released to drive an extinction wave. A one-equation model was then studied for its traveling wave solution, assuming a constant spatial density of sterile insects.

In \cite{qu2019generating}, Qu and Hyman simplified a detailed 9-ODE model \cite{qu2018modeling} into reduced systems of 7, 4, and 2 ODEs, while still capturing important biological dynamics like the basic reproductive number, bifurcations, and threshold conditions. These reduced models retain parameters expressed in terms of the original biologically meaningful variables. Later, they extended these results in \cite{qu2022modeling} by deriving and analyzing a 2-PDE model for \textit{Wolbachia} invasion with spatial dynamics based on the reduced 2-ODE model. We extend this 2-PDE spatial model \cite{Qu_Wu_Hyman_2022} to guide field trials for establishing \textit{Wolbachia} infection among wild \textit{Aedes aegypti} population. 

The results of our model simulations allowed us to identify strategies for optimizing \textit{Wolbachia} releases under constraints on both release size and insecticide efficacy for pre-release interventions. When release size is restricted, \textit{Wolbachia} establishment dynamics are strongly influenced by local mosquito dispersal patterns. In low-dispersal environments, where \textit{Wolbachia}-infected mosquitoes remain concentrated after release, insecticide use may not be necessary to ensure establishment. However, in high-dispersal regions, where mosquitoes spread quickly and may fall below the persistence threshold, pre-release interventions that reduce at least $35\%$ of the wild mosquito population significantly enhance \textit{Wolbachia} establishment within nine months. When insecticide efficacy is limited to $35\%$, release sizes smaller than the constrained value still achieve $90\%$ \textit{Wolbachia} prevalence in low-dispersal areas. However, larger releases are required in high-dispersal regions. Additionally, our results indicate that distributing \textit{Wolbachia} releases in 2–5 weekly batches is more effective than a single large release, even when pre-release insecticide interventions are not feasible. These findings highlight the importance of tailoring both pre-release interventions and release strategies based on local mosquito dispersal features, providing actionable insights for cost-effective and efficient \textit{Wolbachia}-based vector control programs.

The paper is organized as follows. Section \ref{Methods} reviews the transmission dynamics of the chosen spatial model and discusses the integration of physically meaningful parameters. It includes a numerical characterization of the threshold introduction size required to trigger a wave of infection and outlines the setup and rationale for designing various mosquito release scenarios. Section \ref{Results} presents the simulation results, examining the effects of release size, frequency, mosquito dispersion rates, and pre-release interventions such as thermal fogging targeting adult mosquitoes. Finally, Section \ref{Conclusions-Discussion} situates the findings within the broader literature on mosquito-borne disease control, discussing their implications and offering recommendations for future research and policy implementation. This structure provides a comprehensive framework to guide \textit{Wolbachia}-based control strategies.

\section{Methods}\label{Methods}

We base our spatial model on two ordinary differential equations (2-ODE model) and propose a system of two partial differential equations (2-PDE model).  The previous 2-ODE model derived from a detailed 9-ODE model \cite{qu2019generating} that captures the complex transmission of the \textit{w}Mel strain of \textit{Wolbachia} infection among $Aedes~aegypti$ mosquitoes. This 2-ODE model preserves the nonlinear growth terms representing the maternal transmission of \textit{Wolbachia} in the original 9-ODE model. Here, for simplicity, we work with the proposed 2-PDE model in \cite{Qu_Wu_Hyman_2022} to account for mosquito dispersion patterns in two dimensions while reducing computational complexity. 

It is worth mentioning that this 2-PDE system presents backward bifurcation dynamics that are inherited from previous ODE models \cite{Qu_Wu_Hyman_2022}. This threshold imposes the conditions on the number of \textit{Wolbachia}-infected mosquitoes that need to be introduced to establish infection. The estimation for the threshold is for an ideally controlled situation where infected and uninfected cohorts are homogeneously mixed. However, the environmental variation, the wind, and the flight pattern of the infected mosquitoes released can lead to spatial differences in the fraction of infection. Additionally, it is possible that while the infection level exceeds the threshold near the release site, it may be lower near the edges. This argument supports the need to extend previous ODE models to PDE models incorporating diffusion effects \cite{takahashi2005mathematical}, which will account for the spatial heterogeneity of mosquito populations and more accurately simulate the random, unidirectional movement of mosquito flights when they search for food and resources.

In the following subsections, we describe the model dynamics and incorporate model parameters defined in terms of physically meaningful quantities. This definition extends the applicability of our model to provide clear, actionable insights for field practitioners.

\subsection{Review of the 2-PDE \textit{Wolbachia} Transmission Model and its Assumptions}
The foraging behavior of adult female \textit{Aedes aegypti} consists of local flights characterized by a random unidirectional movement. A diffusion process can approximate this phenomenon \cite{takahashi2005mathematical}. We consider the reaction-diffusion PDE system proposed in \cite{Qu_Wu_Hyman_2022} for modeling the spatial dynamics of the vertical transmission of \textit{Wolbachia}, which depends on the infection status of both male and female mosquitoes \cite{walker2011w}. We define $F_u(\mathbf{x},t)$ and $F_w(\mathbf{x},t)$ to be the population density (mosquitoes per square meter) of the adult uninfected and \textit{Wolbachia}-infected mosquitoes at location $\mathbf{x}=(x,y)$ at time $t$. Our 2-PDE model for the birth, death, and diffusion of female mosquitoes is
\begin{subequations}
\begin{align}
\frac{\partial F_u}{\partial t} &= B_u(F_u,F_w) F_u-\mu_{fu}^{r}F_u +D\Delta F_u,\label{eq: 2-PDE-system1-not-scaledKf}\\
\frac{\partial F_w}{\partial t} &= B_w(F_u,F_w)F_w - \mu_{fw}^{r}F_w + D\Delta F_w \label{eq: 2-PDE-system2-not-scaledKf}~.
\end{align}\label{eq: 2-PDE-system1-not-scaledKf-all}
\end{subequations}
We assume a constant uniform diffusion coefficient, $D$, and $\Delta = \partial^2/\partial x^2 + \partial^2/\partial y^2$ is the two-dimensional Laplacian operator. One major assumption in our approach is that we only track adult mosquitoes, a consequence of the model reduction process from the 9-ODE to the 2-ODE framework. This leads to the caveat that the model may not predict field trials that disrupt the natural balance among different life stages or the sexual ratio of the mosquitoes \cite{Qu_Wu_Hyman_2022}. We ameliorate this by considering releases of both male and female mosquitoes in a 1:1 ratio. Throughout the remainder of the paper, any specified quantity of female mosquitoes for the release will also imply the release of approximately an equal number of males.  The complete description of the biologically relevant parameters for the model and their corresponding values and dimensions are presented in Table \ref{tab: ParameterTable-Regular-Model}.
\begin{table}[H]
\begin{adjustwidth}{-2.25in}{0in}
\centering
\small
\caption{
{\textbf{Model parameters, dimensions, and baseline values. Parameter values were retrieved from \cite{Qu_Wu_Hyman_2022}}. Note that the 2-PDE parameters are all defined in terms of the physically meaningful 9-ODE model parameters \cite{qu2018modeling}. Their baseline values differ because the 2-equation model parameters account for the missing states, such as the aquatic stage. For the ODE model, carrying capacity is based on the total number of larvae, while in the PDE model, it is expressed as the density of mosquitoes per $m^2$. A comprehensive overview of the parameters and their values is available in \cite{qu2018modeling}.\vspace{0.1in}}}
\begin{tabular}{|l|l|l|l|}
\hline
\multicolumn{4}{|l|}{\bf Biologically Relevant Parameters (9-ODE)} \\ \thickhline
\textbf{Parameter} & \textbf{Description} & \textbf{Baseline} & \textbf{Dimensions} \\ \hline
$b_f$ & Female birth probability & 0.5 & - \\ \hline
$\nu_w$ & Maternal \textit{Wolbachia} transmission probability & 1 & - \\ \hline
$\nu_u$ & $1-\nu_w$ & 0 & - \\ \hline
$\sigma$ & Per capita mating rate & 1 & day$^{-1}$ \\ \hline
$\phi_u$ & Per capita egg-laying rate for uninfected females & 13 & eggs/day \\ \hline
$\phi_w$ & Per capita egg-laying rate for infected females & 11 & eggs/day \\ \hline
$\psi$ & Per capita egg development rate & $1/8.75$ & day$^{-1}$ \\ \hline
$\mu_a$ & Death rate for aquatic stage adults & 1/50 & day$^{-1}$ \\ \hline
$\mu_{fu}$ & Death rate for uninfected females & $1/17.5$ & day$^{-1}$ \\ \hline
$\mu_{fw}$ & Death rate for infected females & $1/15.8$ & day$^{-1}$ \\ \hline
$K_a$ & Carrying capacity of aquatic stage & - & \# of larvae \\ \hline
\multicolumn{4}{|l|}{\bf Adjusted Parameters of Reduced 2-PDE Model} \\ \thickhline
$\phi_u^{r}$ & Adjusted per capita reproduction rate for $F_u$ & 7 & day$^{-1}$ \\ \hline
$\phi_w^{r}$ & Adjusted per capita reproduction rate for $F_w$ & 5.7 & day$^{-1}$ \\ \hline
$\mu_{fu}^{r}$ & Adjusted death rate for $F_u$ & $1/26.25$ & day$^{-1}$ \\ \hline
$\mu_{fw}^{r}$ & Adjusted death rate for $F_w$ & $1/24.55$ & day$^{-1}$ \\ \hline
$K_f$ & Carrying capacity for females & 3 & density of mosquitoes/$m^2$ \\ \hline
$D$ & Diffusion coefficient & {[}10-100{]} & $m^2$/day \\ \hline
\end{tabular}
\label{tab: ParameterTable-Regular-Model}
\end{adjustwidth}
\end{table}

The terms $B_u$ and $B_w$ in the system of equations \ref{eq: 2-PDE-system1-not-scaledKf-all} represent the per capita birth functions for uninfected and infected mosquitoes, respectively. The parameters $\mu_{fu}$ and $\mu_{fw}$ are their death rates, while $K_f$ indicates the carrying capacity of the adult female population. These birth functions are defined as:
\begin{subequations}
\begin{align}
B_u(F_u,F_w)&=  b_f\phi_u^{r}\frac{\mu_{fu}^{r}F_u}{\mu_{fu}^{r}F_u + \mu_{fw}^{r}F_w}\Bigg(1-\frac{F_u+F_w}{K_f}\Bigg), \label{eq: birth-rate_u}\\
B_w(F_u,F_w) &=  b_f\phi_w^{r}\Bigg(1-\frac{F_u+F_w}{K_f}\Bigg). \label{eq: birth-rate_w}
\end{align}\label{eq: birth-rate_all}
\end{subequations}

Note that each birth rate function is written as a product of different factors representing the mosquito birth cycle based on infection status. The first one, $b_f\phi_u^{r}$ in  Eq \ref{eq: birth-rate_u} and $b_f\phi_w^{r}$ Eq \ref{eq: birth-rate_w}, corresponds to the per capita reproduction rates of uninfected and infected females, respectively. Here, we assume that half of the offspring will develop into the next generation of females ($b_f = 1/2$).
Due to the perfect maternal transmission assumption of \textit{Wolbachia} in our model, all the offspring produced 
by the infected females is \textit{Wolbachia}-infected regardless of the infectious status of the males \cite{qu2022modeling}.

It is worth highlighting that the strain \textit{w}Mel of \textit{Wolbachia} we are working with exhibits a strong cytoplasmic incompatibility (CI) phenomenon \cite{walker2011w}. This means that when an infected male mates with an uninfected female, no viable offspring are produced \cite{laven1967eradication}, this phenomenon is incorporated in the term $B_u$ of Eq. \ref{eq: birth-rate_u}.
 
Based on the equation reduction process derived in equation 3.11 in \cite{qu2019generating} to reduce the number of independent variables, 
the second factor in Eq. \ref{eq: birth-rate_u}, $\mu_{fu}^{r}F_u/(\mu_{fu}^{r}F_u+\mu_{fw}^{r}F_w)$, corresponds to an approximation of the fraction of uninfected males among total males, $M_u/(M_u+M_w)$. Here, $M_u$ and $M_w$ are the populations for the uninfected and infected males, respectively, 
\cite{qu2019generating}. Note that if the death rates of the infected and uninfected mosquitoes were the same, $\mu_{fu}^{r}=\mu_{fw}^{r}$, then the ratio of uninfected females and the ratio of uninfected males would be the same.  This approximation is biologically intuitive, as it relies on the simplifying yet realistic assumption that an equal number of male and female mosquitoes are born, with the ratio between them only influenced by the differences in their death rates. 

All the birth rate terms are regularized by the female carrying capacity, $K_f$. More precisely, we incorporate the logistic growth model term $(1-\frac{F_u+F_w}{K_f})$ in both birth functions of Eqs. \ref{eq: birth-rate_all}. This corresponds to a regularization of the total adult female population and can be interpreted as limiting the availability of breeding sites for adult females \cite{rejmankova2013ecology}.

In our framework, we rescale the \textit{Wolbachia} models state variables relative to the female carrying capacity $K_f$. This rescaling enhances the model adaptability, enabling results to be applied to regions with varying ecological conditions and population sizes, thereby supporting the broader implementation of \textit{Wolbachia}-based intervention strategies in diverse settings. With this in mind, if we define $u=F_u/K_f$ and $w=F_w/K_f$ as the \textit{Wolbachia} uninfected and infected mosquito population sizes, respectively, expressed as fractions of the female carrying capacity, our equations above will be given by:
\begin{subequations}
\begin{align}
\frac{\partial u}{\partial t} &= b_f\phi_u^{r}\frac{\mu_{fu}^{r}u}{\mu_{fu}^{r}u + \mu_{fw}^{r} w}(1-u-w)u-\mu_{fu}^{r}u +D\Delta u,\label{eq: 2-PDE-system1-scaledKf}\\
\frac{\partial w}{\partial t} &= b_f\phi_w^{r}(1-u-w)w - \mu_{fw}^{r}w + D\Delta w. \label{eq: 2-PDE-system2-scaledKf}
\end{align}
\end{subequations}

It is worth highlighting that the diffusion coefficient $D$ measures the mean absolute deviation of the mosquito flights per day \cite{takahashi2005mathematical}. This additional measurement corresponds to the spatial extension Qu and Hyman proposed in \cite{Qu_Wu_Hyman_2022} to their previous 2-ODE model \cite{qu2019generating}. This extension preserved the backward bifurcation dynamics on the 2-ODE system, identifying a critical threshold condition for invasion. More precisely, assuming perfect maternal transmission, the system has a stable \textit{Wolbachia}-free equilibrium point (WFE), a stable complete infection equilibrium (CIE), and an unstable threshold endemic equilibrium (EE) where both infected and uninfected cohorts coexist. When the infection level is above the threshold, the infection eventually takes off and approaches the CIE; when below this level, the system approaches the WFE, and any minor releases of infected mosquitoes will die out \cite{qu2022modeling}.

The reason behind using two spatial dimensions in our model relies entirely on the need to guide field releases of \textit{Wolbachia} infected female mosquitoes. These releases are typically done in a local region or point. The resulting infection wave propagates outward as a radial expansion. With this in mind, we simplify the model by assuming that the effect of wind and environmental heterogeneity is minimal and that the spread can be approximated by assuming cylindrical symmetry. Given a coordinate point in the 2D Cartesian space $(x,y)$, we consider the change of coordinates, $r=\sqrt{x^2+y^2}$, to account for this radial expansion effect. This leads to the following cylindrically symmetric system of equations:

\begin{subequations}
\begin{align}
\frac{\partial u}{\partial t} &= b_f\phi_u^{r}\frac{u}{u + \frac{\mu_{fw}^{r}}{\mu_{fu}^{r}}w}(1-u-w)u-\mu_{fu}^{r}u +D\Bigg(\frac{\partial u^2}{\partial r^2}+ \frac{1}{r}\frac{\partial u}{\partial r}\Bigg),\label{eq: 2-PDE-system1-cyllindrical}\\
\frac{\partial w}{\partial t} &= b_f\phi_w^{r}(1-u-w)w - \mu_{fw}^{r}w + D\Bigg(\frac{\partial w^2}{\partial r^2}+ \frac{1}{r}\frac{\partial w}{\partial r}\Bigg). \label{eq: 2-PDE-system2-cyllindrical}
\end{align}
\end{subequations}
The derivation of the system and the corresponding numerical method for approximating the solution can be found in Appendices \ref{cyllindrically-symmetric-derivation}-\ref{numerical-method}, respectively.

In the following section, we use the mathematical framework previously defined to determine the critical threshold conditions, i.e., the minimal release size of infection, in two spatial dimensions to establish a wave invasion of \textit{Wolbachia} near the release center \cite{Qu_Wu_Hyman_2022}.

\subsection{Characterization of a Threshold of Infection}\label{chatacterize-threshold}

\textit{Wolbachia} deployment trials require consistent monitoring to ensure infected mosquito populations' establishment and long-term sustainability. Key monitoring strategies include deploying mosquito traps and regular molecular testing to assess \textit{Wolbachia} presence. Typically, these trials rely on specific infection frequency thresholds that must be achieved and sustained over several weeks before ceasing releases. For instance, deployments in Rio de Janeiro utilized BG-Sentinel traps to monitor infection rates weekly during the initial release phases and at intervals of 1–3 months after establishment \cite{durovni2020impact}.

In what follows, we propose a computational framework to identify a threshold—the minimum quantity of \textit{Wolbachia}-infected mosquitoes required for introduction into a 2D spatial domain—to ensure that the infection frequency remains at or above $60\%$ for at least five years. This approach simplifies operational efforts by alleviating the need to monitor weekly infection rate targets continuously. Moreover, it reduces overall monitoring costs by decreasing the frequency and intensity of labor-intensive tasks such as trap deployment, morphological identification, and molecular testing \cite{durovni2020impact, velez2023large}. By establishing a stable threshold, this method makes \textit{Wolbachia} deployments more cost-effective and scalable across diverse ecological and urban environments.
 
The threshold condition determines a minimum quantity of \textit{Wolbachia}-infected mosquitoes that will create a sustained infection in the field. The analytical results presented in \cite{Qu_Wu_Hyman_2022} and \cite{fife2013mathematical} suggest that the ODE threshold corresponds to a PDE threshold when extended to the spatially homogeneous setting. However, this poses a practical drawback for guiding field releases since it requires a positive infection on an infinite domain. 

For the scope of our simulations, a suitable threshold condition for a realistic field release of infected mosquitoes requires a release over a concentrated region. This type of release is known as point release and allows the model to simulate mosquito dynamics more realistically. In particular, for simulating natural mosquito dispersal dynamics \cite{russell2005mark}, since mosquitoes tend to concentrate around resources such as water, shade, and human habitats \cite{winskill2015dispersal}. To characterize the infection threshold, Qu and Hyman in \cite{qu2022modeling} used a point-release process that eventually transitioned into a critical, bubble-shaped infection profile. Here, to approximate the geometry of this critical bubble, we will use an inverse squared exponential distribution as an initial smooth release profile of infected mosquitoes. We also assume that the uninfected mosquito population starts at carrying capacity level, i.e., $K_f=3$ mosquitoes per $m^2$. The corresponding formulations for the initial conditions where the uninfected wild mosquitoes are at the fraction $u_o$ of the carrying capacity and the \textit{Wolbachia}-infected mosquitoes are released as Gaussian function, 
\begin{subequations}
\begin{align}
u(r,0)&=u_o, \label{eq: initial_cond_u}\\
w(r,0)&=\frac{W}{2\pi K_f\sigma_{w}^2}e^{-r^2/2\sigma_{w}^2}.\label{eq: initial_cond_v}
\end{align}
\end{subequations}

Unless stated otherwise, we assume that the wild mosquitoes are initially at the carrying capacity, $u_o=1$. This assumption considers an initial value close to but not equal to the disease-free equilibrium of the system of equations, whose derivation can be found in \cite{qu2022modeling}.
Here, $W$ corresponds to the total number of \textit{Wolbachia} infected mosquitoes that will be introduced in our domain, $K_f$ is the carrying capacity, and $\sigma_w$ is the standard deviation of the Gaussian release.

\begin{figure}[hbtp]
\centering
\includegraphics[width=.85\linewidth]{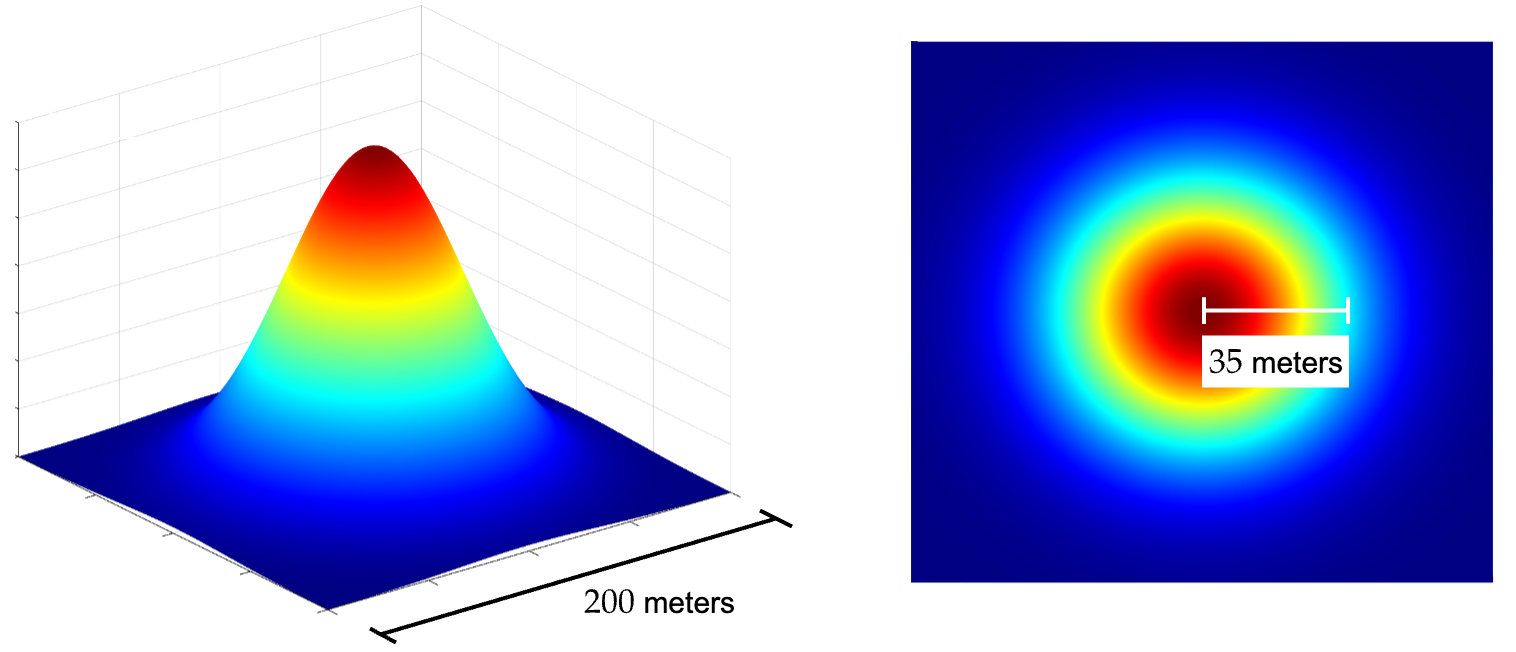}
\caption{\small \textbf{Release shape geometry}  Graphical representation of the squared exponential release over a 2D domain size of $3000\;meters\times3000\;meters$. The left plot highlights the release over a subset of the domain ($200\;meters\times200\;meters$), while the right plot illustrates the Gaussian release standard deviation of $\sigma_W=35$ meters with respect to the domain center.}
\label{fig: graphical_release_shape}
\end{figure}

Given this initial shape profile, we can now identify the corresponding threshold condition parameterized by its infection level at the release center and with a standard deviation of $\sigma_w=35$ meters with respect to this center. Our choice for $\sigma_w$ focuses on representing short-range mosquito dispersal expected in urban environments \cite{filipovic2020using}.

We define $p_{thres}$ to be the threshold condition for \textit{Wolbachia} infection establishment. To compute it, we consider the following formula that characterizes the level of infection at the peak at time $t$, in terms of the state variables,
\begin{equation}
p_{peak}(t) = \frac{w(r_0,t)}{u(r_0,t)+w(r_0,t)},\;\; \text{for}\;\; t>0, r_0=0. \label{eq: p_peak-parametrization}
\end{equation}

The previous formula should be interpreted as the fraction of \textit{Wolbachia} infected mosquitos at the release center $r_0=0$ for a given time $t$. 
We employ the bisection method \cite{sikorski1982bisection} on initial infection values and keep track of their peak of infection $p_{peak}$ for a sufficiently long time (about five years $\approx$ 2000 days).
\begin{figure}[hbtp]
\centering
\begin{minipage}{.5\textwidth}
  \centering
  \includegraphics[width=0.99\linewidth]{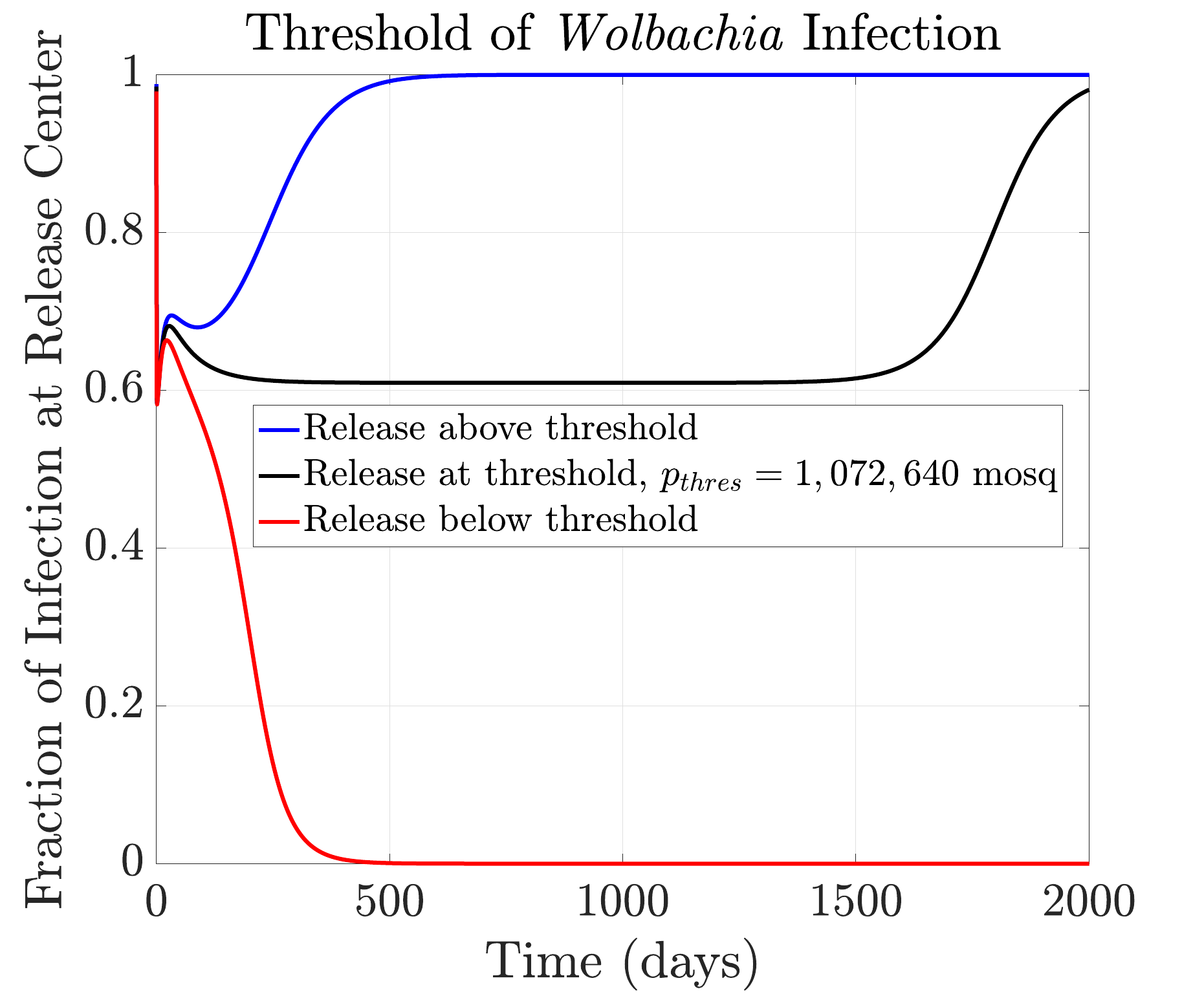}
\end{minipage}%
\begin{minipage}{.5\textwidth}
  \centering
  \includegraphics[width=1.05\linewidth]{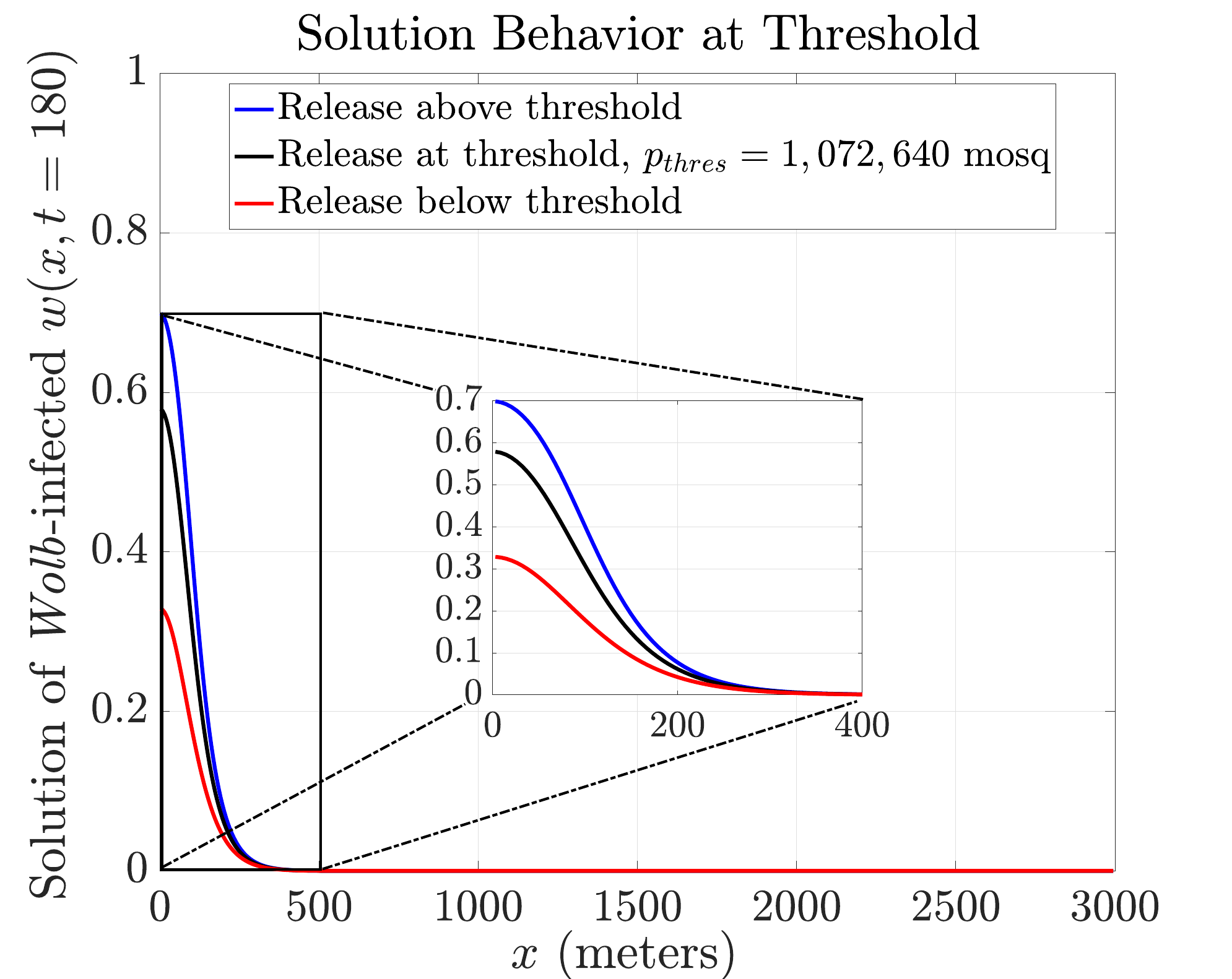}
\end{minipage}
\caption{\small\textbf{Numerical computation of threshold condition (PDE solver time step $\Delta t=1.78\times 10^{-2}$, diffusion coefficient $D=60\;m^2/day$)}. \textbf{Left}: fraction of infection curves for different initial release sizes: releasing at the threshold, i.e,  $p_{\text{thres}} = 1,072,640$ female mosquitoes (assuming same amount of male mosquitoes are also released), which is approximately $35.7\%$ of the female carrying capacity (black curve), above the threshold (blue curve), and below the threshold (red curve). After a transition period of approximately 250 days, the infection curve at the threshold stabilizes at a level of $0.6$ over an extended period. In contrast, the curves for releases above or below the threshold quickly approach the complete infection or no-infection states, respectively. \textbf{Right}: behavior of the PDE solution for the \textit{Wolbachia}-infected state variable $w(r,t)$ (at $t=180$ days) when releasing \textit{Wolbachia}-infected mosquitoes at threshold value (black curve), above threshold (blue curve), and below threshold (red curve). The solution curve at the threshold evolves into a bubble shape at day $180$ and preserves this shape for about four years.}  
\label{fig: Threshold-computation}
\end{figure}
The threshold condition, $p_{thres}$, picked in this case, is the minimum initial  \textit{Wolbachia} release quantity, obtained by integrating the initial condition formula in Eq. \ref{eq: initial_cond_v}, that leads to the curve that stays at a constant level of infection for a longer time, and that satisfies
\begin{equation}
0.9 <p_{peak}(t_{2000}) < 0.9 + \epsilon. \label{eq: condition-theshold-computation}
\end{equation} 

The value that satisfies the previous condition is $p_{thres}=1,072,640$ female and male mosquitoes, which corresponds to approximately $35.7\%$ of the female carrying capacity $K_f=3,000,000$ (per $km^2$). This analysis was performed with a time step of the PDE solver of $\Delta t=1.78\times10^{-2}$ days and a Gaussian standard deviation of $35$ meters with respect to the release center. It is worth highlighting that the numerical resolution of the threshold depends on the time step chosen for the PDE solver. While the threshold condition may vary slightly with different time steps, our numerical convergence test (detailed in Appendix \ref{convergence-test-threshold}) confirms our results do not suffer from numerical artifacts and accurately reflect the model dynamics.

The analysis result is depicted in Figure \ref{fig: Threshold-computation}. Here, we present the behavior of both the fraction of infection curves at the release center (left panel) and the PDE solution (right panel) when releasing mosquitoes at three different release sizes: at threshold level ($p_{thres}=1,072,640$ female mosquitoes, assuming an equal number of males are released), equivalent to $35.7\%$ of the female carrying capacity), represented by the black curve, above the threshold (blue curve), and below the threshold (red curve). 

In Figure \ref{fig: Threshold-computation} (left), we observe that if we release infected mosquitoes below the threshold (red curve), the fraction of infection will decrease until infection dies out. If we release above that threshold, the curve will reach complete infection quickly. But if we release \textit{Wolbachia}-infected mosquitoes at threshold level, the 
 fraction of the infection curve reaches a steady level of approximately $0.6$ for about four years before eventually establishing completely among the population.

When releasing the infected mosquitoes at the threshold value, the solution, shown in Figure \ref{fig: Threshold-computation} (right), converges to a bubble-shaped profile after a transition period of approximately $180$ days. This shape balances the competition between the growth of infection from reproduction of \textit{Wolbachia} infected females in the reaction term versus the spread of infection from mosquito diffusion.

For the scope of the simulations of field release trials, we aim to release an infection level above the threshold condition profile to establish the wave of \textit{Wolbachia}. This threshold condition that leads to a critical bubble serves as a theoretical reference for wave initiation. However, it may not be an ideal release design if a faster establishment is required. For this reason, to inform a more practical field release in the upcoming simulations, we consider a target infection level of $90\%$ of \textit{Wolbachia} infected mosquitoes to be achieved within one year after doing an initial release above the critical threshold.

\subsection{Design of Release Scenarios}

 In our simulations, we consider releasing \textit{Wolbachia} infected mosquitoes over an urban area. Evidence suggests that mosquitoes of the \textit{Aedes aegypti} species prefer to live near their breeding site \cite{longdistancesCDC}. For this reason, although the scope of their flight range can reach as far as 850 meters \cite{reiter1995dispersal}, during their lifetime, they will only fly a few blocks \cite{longdistancesCDC}. Thus, in our simulations, we consider a diffusion coefficient range between $10-100\;m^2/$day to represent mosquito dispersal of this species.
 
 For the upcoming release scenarios, we assume to release the same amount of \textit{Wolbachia} infected males and females (ratio 1:1, females:males). Although the current model only accounts for female mosquito populations, with the model reduction process derived in \cite{qu2019generating}, we assume that approximately the same number of male mosquitoes are released.

Building upon these assumptions, we evaluate strategies for improving the establishment of a wave of infection. Specifically, we assess the effects of infection release size, repetitive releases, and insecticide-based pre-release strategies. These strategies are motivated by challenges observed during large-scale releases where \textit{w}Mel failed to establish fully. For instance, in Medellín, Colombia, \textit{Wolbachia} prevalence declined to $33.5\%$ two years after a release program (2017–2022), with significant variation across urban communes \cite{calle2024evaluation}. Similarly, in Rio de Janeiro, Brazil, releases conducted between 2017 and 2019 resulted in only $32\%$ prevalence after 29 months, despite large release efforts \cite{dos2022estimating}.

These examples highlight the challenges of achieving consistent success in urban environments, particularly when resources for sustained releases or monitoring are limited. To address these issues, we explore the potential of combining \textit{Wolbachia} release programs with complementary intervention strategies, such as repetitive releases and pre-release insecticide spraying, to mitigate the re-invasion of uninfected mosquito populations and improve the long-term establishment of \textit{Wolbachia}.

To enhance infection establishment and minimize the number of infected mosquitoes required for release, a common approach involves reducing the uninfected mosquito population through insecticide spraying before the release of \textit{Wolbachia}-infected mosquitoes. This method, often called pre-release insecticide application, involves dispersing a liquid fog of insecticide in outdoor areas to target adult mosquitoes directly. In this study, we will refer to this process as thermal fogging. This strategy has been shown to enhance the effectiveness of \textit{Wolbachia}-based disease control \cite{qu2018modeling, hu2021mosquito, hoffmann2013facilitating}.

In addition to insecticide-based pre-release measures, repetitive releases, where infected mosquitoes are released periodically in smaller batches, are widely used in field trials \cite{zheng2021one}. While previous studies have compared single large releases to multiple smaller releases \cite{hancock2011strategies, florez2023modeling}, they often overlook spatial heterogeneity in the release area, which can significantly influence the outcomes.

Using a spatially explicit model, we aim to bridge this gap by investigating how spatial variability affects the efficacy of these strategies. Specifically, we explore the impact of insecticide-based pre-release measures and phased releases on the proportion of \textit{Wolbachia}-infected mosquitoes required to establish infection across regions with different mosquito dispersal patterns.

\subsection{Incorporating Pre-release Strategies and Repetitive Releases}
Thermal fogging, as a pre-release strategy, is implemented to reduce the uninfected mosquito population prior to the release of \textit{Wolbachia}-infected mosquitoes. Since the model described in Eq. \ref{eq: 2-PDE-system1-cyllindrical}-\ref{eq: 2-PDE-system2-cyllindrical} only includes adult mosquito compartments, we assume that thermal fogging involves the application of insecticide fog to target adult mosquitoes directly.

To incorporate multiple levels of thermal fogging intensity, we adjust the initial conditions of our system of PDEs (Eq. \ref{eq: initial_cond_u}-\ref{eq: initial_cond_v}) for the uninfected population. More precisely,  the uninfected female population $F_u(r,t)$ satisfies $0\leq F_u(r,t)\leq K_f$, where $K_f$ is the female carrying capacity. When scaled by $K_f$, this inequality can be written as $0\leq u(r,t)\leq 1$.

Now, let us define $\alpha$ as the fraction of the carrying capacity to be removed from the population using thermal fogging. The initial conditions for the uninfected females are defined with a fraction of mosquitoes of $u(r,0)=1-\alpha$, adjusted from the carrying capacity. 

Regarding the implementation of phased releases, we compare the effect of releasing the total amount \textit{Wolbachia}-infected mosquitoes distributed in periodic weekly batches versus releasing them all at once. Since the average lifespan of an adult \textit{Aedes aegypti} mosquitoes ranges between 2-4 weeks \cite{goindin2015parity}, we consider repetitive releases ranging between 1-5 weeks for our simulations. This ensures that at least one full generation of wild mosquitoes is exposed to \textit{Wolbachia} infection \cite{zheng2021one,hoffmann2011successful}. This overlap might increase the likelihood of \textit{Wolbachia } spreading within the population through mating.

\section{Results}\label{Results}
Based on the model assumptions and the simulation design described previously, we present the results, emphasizing the best achievable outcomes under resource limitations on: 1) the number of \textit{Wolbachia}-infected mosquitoes that can be introduced through a deployment program, and 2) The efficacy level of insecticide used to reduce the uninfected mosquito population before the first release.  Additionally, we evaluate the benefits of releasing infected mosquitoes in weekly batches compared to a single mass release, considering these two resource limitations.

\subsection{Model Results for a Fixed Release Size of \textit{Wolbachia} Infected Mosquitoes}

\begin{figure}[hbtp]
\centering
\includegraphics[width=1\linewidth]{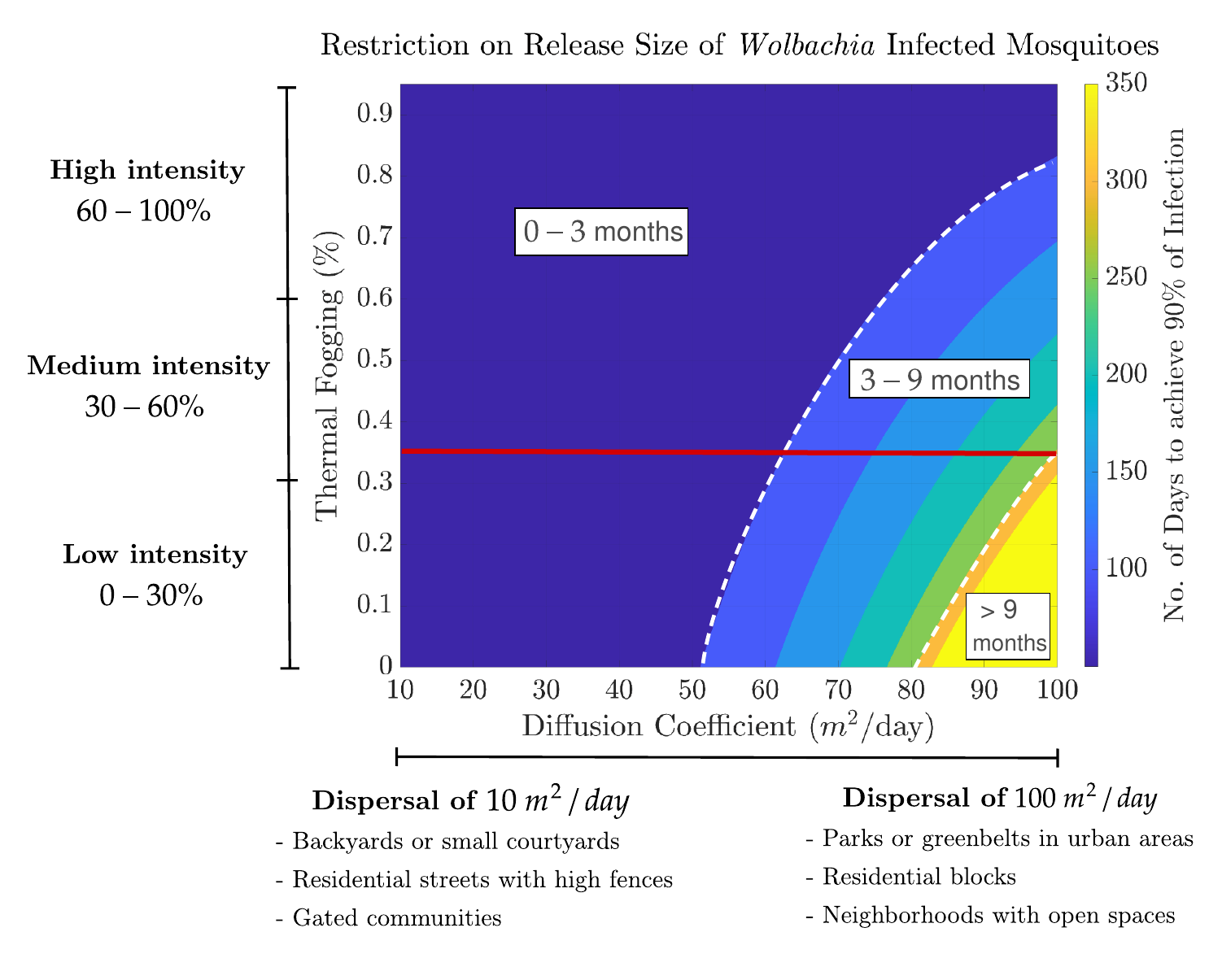}
\caption{\small Heatmap showing the time in months (region labels) and days (right y-axis), required to achieve $90\%$ of \textit{Wolbachia} infection in a mosquito population,  based on a fixed release size of $3,600,000$ female infected mosquitoes (and about the same amount of males) in a $3000m\times3000m$ area. The $x$-axis represents the diffusion coefficient in $m^2/day$, reflecting mosquito dispersal rates, while the $y$-axis indicates the proportion of the wild mosquito population removed through thermal fogging with three different ranges of intensities. The heatmap color code corresponds to the time to reach $90\%$ of infection, with darker shades indicating shorter times. The figure highlights three distinct regions: $0-3$ months (dark), $3-9$ months (intermediate), and more than $9$ months (light). Removing at least $35\%$ (red line) of the uninfected mosquito population ensures $90\%$ infection is reached within nine months, regardless of mosquito dispersal, while lower removal levels lead to longer times, particularly at large diffusion coefficients.}
\label{fig: S1-fixed-release-pre-release}
\end{figure}

The time required to achieve $90\%$ \textit{Wolbachia} infection in a mosquito population depends on mosquito dispersal rates and the intensity of pre-release interventions, such as thermal fogging. Figure \ref{fig: S1-fixed-release-pre-release} explores this relationship under a resource-constrained release of $3,600,000$ \textit{Wolbachia}-infected female mosquitoes (and an equal number of males) into a $3,000\;m\times3,000\;m$ area, which is comparable to the field trial deployed in the municipality of Bel\'en in Medell\'in, Colombia \cite{velez2023large}. By holding the release size constant, the analysis isolates the effects of varying mosquito dispersal rates and pre-release reductions of uninfected mosquitoes on infection establishment. Additionally, the diffusion coefficient range used to represent mosquito dispersal rates is contextualized with descriptions of the landscape features associated with the lowest dispersal rate ($10\;m^2/day$) and the highest dispersal rate ($100\;m^2/day$) considered ($x-$axis in Figure \ref{fig: S1-fixed-release-pre-release}).

The heatmap in Figure \ref{fig: S1-fixed-release-pre-release} illustrates three distinct temporal regions for infection establishment. The dark region indicates establishment between $0-3$ months. In this region, low dispersal rates $10-50\; m^2/day$ help achieve $90\%$ of infection within three months regardless of the intensity of thermal fogging applied. This can be attributed to the interplay between mosquito movement and the dynamics of infection spread. More precisely, at low dispersal rates, infected mosquitoes released into the environment are less likely to disperse widely, leading to a higher concentration of infected individuals near the release site. This clustering effect increases the likelihood of local mating of the released mosquitoes, accelerating the spread of \textit{Wolbachia}.

The reduced reliance on thermal fogging in this region may also stem from the limited overlap between infected and uninfected mosquitoes, meaning that even moderate natural population dynamics (e.g., birth and death) combined with infection spread dynamics suffice to establish high infection rates without requiring intense intervention. Previous studies have also observed this phenomenon of rapid \textit{Wolbachia} establishment at low dispersal rates due to localized interactions, such as, \cite{hoffmann2011successful, schmidt2017local}.

As the diffusion coefficient increases to dispersal rates of $50\;m^2/day$ or higher, achieving $90\%$ \textit{Wolbachia} infection within $3-9$ months (intermediate region in Figure \ref{fig: S1-fixed-release-pre-release}) or beyond $9$ months (light region in Figure \ref{fig: S1-fixed-release-pre-release}) requires incorporating thermal fogging as a supplementary intervention in the release program. It is worth noticing that at dispersal rates between $80-100\;m^2/day$, infection establishment dynamics indicate that a medium intensity of thermal fogging (at least $35\%$ represented by the horizontal red line in Figure \ref{fig: S1-fixed-release-pre-release}) is the minimum required to achieve $90\%$ of infection within $9$ months post-release.

At higher dispersal rates, the released infected mosquitoes mix more rapidly with the uninfected populations over larger areas within the domain, causing infection dilution near the release center and making it harder to establish. Thermal fogging helps by reducing the number of uninfected mosquitoes, increasing the relative proportion of \textit{Wolbachia}-infected individuals in the population and speeding up the spread process. This reduction directly affects the threshold necessary for infection establishment, making achieving it more feasible.

This analysis offers insights for designing release strategies under resource constraints on the number of \textit{Wolbachia}-infected mosquitoes, while accounting for mosquito dispersal rates shaped by the landscape features of the release domain. Specifically, for dispersal rates between $10-50\;m^2/day$, typical of areas with restricted movement such as backyards or gated communities, \textit{Wolbachia} mosquitoes remain concentrated locally after release, which may benefit infection establishment without the use of pre-release insecticides. In contrast, for dispersal rates between $50-100\;m^2/day$, associated with more open environments like residential blocks, parks, or urban greenbelts, incorporating pre-release interventions—removing at least $35\%$ of the uninfected mosquito population—is recommended to enhance infection establishment.

\subsection{Model Results for a Fixed Level of Insecticide Efficacy (Thermal Fogging) Sprayed on Uninfected Mosquito Population}
Based on the minimum level of insecticide efficacy identified in the previous scenario ($35\%$ intensity of thermal fogging), we now focus on a scenario restricting insecticide use to this minimum intensity level. This analysis examines how release size and dispersal rates interact to determine the time required to establish $90\%$ \textit{Wolbachia} infection. By holding the level of insecticide efficacy constant, this analysis isolates the effects of dispersal rates and release size on infection establishment, identifying potential regimes where infection can be established within nine months using fewer mosquitoes than the reference value used in our simulations ($3,600,000$ females and about the same number of males).

\begin{figure}[hbtp]
\centering
\includegraphics[width=1\linewidth]{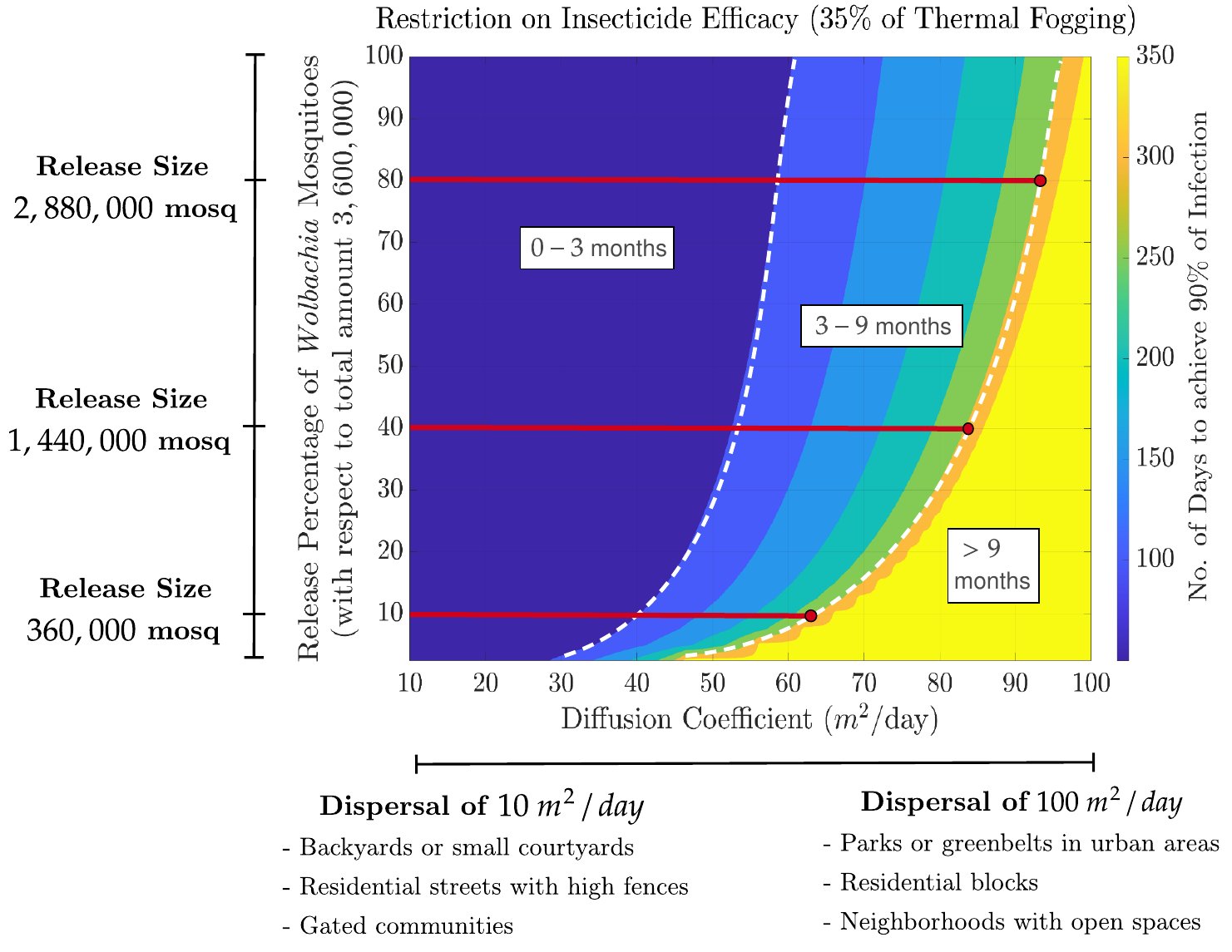}
\caption{\small Heatmap showing the time in months (region labels) and days (right y-axis) required to achieve $90\%$ \textit{Wolbachia} infection in a mosquito population, assuming a restricted level of insecticide efficacy ($35\%$ intensity of thermal fogging) sprayed on uninfected mosquitoes in a $3000\,m \times 3000\,m$ area. The $x$-axis represents the diffusion coefficient in $m^2/\text{day}$, reflecting mosquito dispersal rates. The $y$-axis indicates the percentage of \textit{Wolbachia}-infected mosquitoes to release with respect to the reference baseline value used in the simulations ($3,600,000$ females and about the same amount of males). The heatmap color code corresponds to the time to reach $90\%$ of infection, with darker shades indicating shorter establishment times. The figure highlights three distinct regions: $0$--$3$ months (dark), $3$--$9$ months (intermediate), and more than $9$ months (light). For diffusion coefficients between $10$--$45\,m^2/\text{day}$, even a small quantity of released mosquitoes can establish infection within $0$--$9$ months. For dispersal rates greater than $45\,m^2/\text{day}$, the dashed white contour line indicates the minimum quantity of \textit{Wolbachia}-infected mosquitoes that must be released to reach $90\%$ of infection within $9$ months. The horizontal solid red lines indicate example values of dispersal rates and their corresponding release quantities.}
\label{fig: S2-fixed-pre-release}
\end{figure}

The heatmap in Figure \ref{fig: S2-fixed-pre-release} illustrates three distinct temporal regions for infection establishment under a fixed thermal fogging killing efficacy ($35\%$ intensity) applied to the uninfected population. Infection establishment can occur within $0-3$ months (dark region), between $3-9$ months (intermediate region), or take more than $9$ months (light region). 
The regional differentiation described above allows us to identify two types of infection spread regimes based on dispersal rate ranges (see Figure \ref{fig: S2-fixed-pre-release}). The first regime corresponds to dispersal rates between $10-45\;m^2$/day, where even a small release quantity of \textit{Wolbachia}-infected mosquitoes is sufficient to establish infection within nine months. At these lower dispersal rates, where mosquito movement is more limited, the uninfected and infected cohorts are more likely to encounter each other and mate. This increased likelihood of interaction, combined with the reduced competition for mates and resources due to the pre-release killing of $35\%$ of the uninfected cohort, enhances the reproductive success of the \textit{Wolbachia}-infected mosquitoes via cytoplasmic incompatibility. These factors together explain why a smaller number of \textit{Wolbachia}-infected individuals is needed to establish the infection within the population successfully \cite{de2017does}.

The second regime corresponds to diffusion coefficients greater than \(45\;m^2/\text{day}\), where mosquito dispersal is higher, leading to a more extensive spread of both infected and uninfected individuals. In this case, the dashed contour line in Figure \ref{fig: S2-fixed-pre-release} represents the minimum number of \textit{Wolbachia}-infected mosquitoes that must be released (relative to the reference value of ($3,600,000$ females) to achieve $90\%$ infection within nine months. 

At higher diffusion rates, mosquitoes disperse over a larger area more quickly, reducing the frequency of encounters between infected and uninfected individuals. This dilution effect diminishes the impact of cytoplasmic incompatibility in suppressing the uninfected population. As a result, the pre-release intervention of killing \(35\%\) of the uninfected mosquitoes is no longer sufficient on its own. To overcome this challenge, the release size of \textit{Wolbachia}-infected mosquitoes must be increased to ensure sufficient contact between cohorts and to maintain the spread of \textit{Wolbachia} throughout the population.

It is worth noticing that it is still possible to achieve infection with release quantities smaller than the reference value ($3,600,000$ for female and male mosquitoes) in this regime. These minimum release values, indicated by the dashed contour line (see Figure \ref{fig: S2-fixed-pre-release}), increase as the dispersal rate rises, reflecting the greater challenge posed by higher mosquito mobility. Notably, infection establishment within nine months becomes unachievable for dispersal rates approaching $100\;m^2$/day, even when releasing the full reference quantity of \textit{Wolbachia}-infected mosquitoes.
This indicates that at very high dispersal rates, the spread of uninfected mosquitoes becomes too rapid for the \textit{Wolbachia}-infected cohort to establish the infection within the desired time frame.

The previous analysis provides recommendations for appropriate release sizes of \textit{Wolbachia}-infected mosquitoes in different dispersal scenarios, considering the resource limitations on the efficacy level of insecticide used for pre-release interventions. Specifically, in areas with limited mosquito dispersal ($10-45\;m^2$/day), such as small courtyards or residential streets with high fences, smaller release quantities of \textit{Wolbachia}-infected mosquitoes would be sufficient for effective infection spread. However, larger release sizes would be necessary to ensure successful infection establishment in areas with higher mosquito dispersal (greater than $45\;m^2$/day). 

\subsection{Effect of Repetitive Releases}\label{mult_references_ch4}
\begin{figure}[hbtp]
\centering
\begin{minipage}{.5\textwidth}
  \centering
  \includegraphics[width=1.03\linewidth]{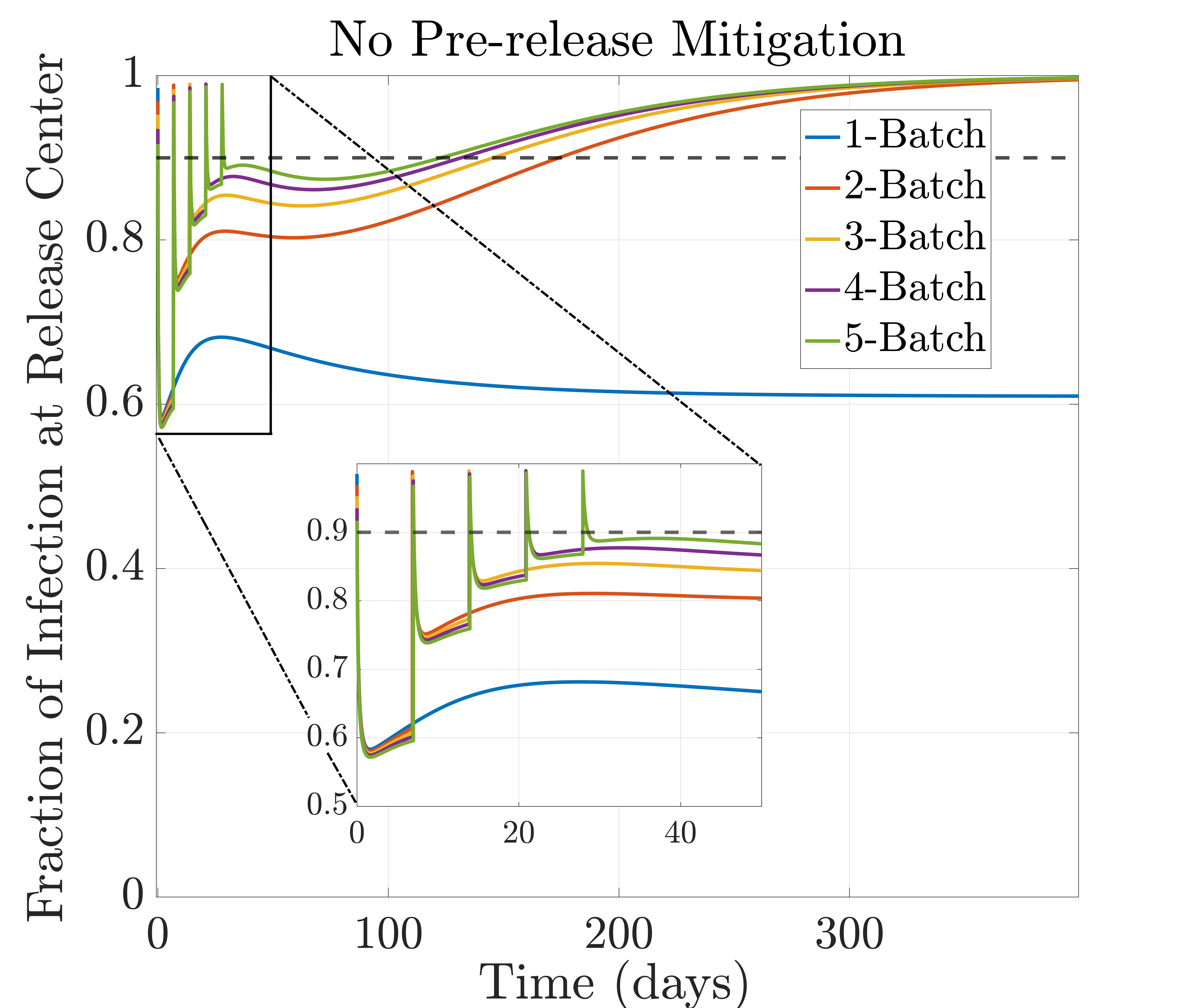}
  \vspace{.05in}
\end{minipage}%
\begin{minipage}{.5\textwidth}
  \centering
  \includegraphics[scale=0.0263]{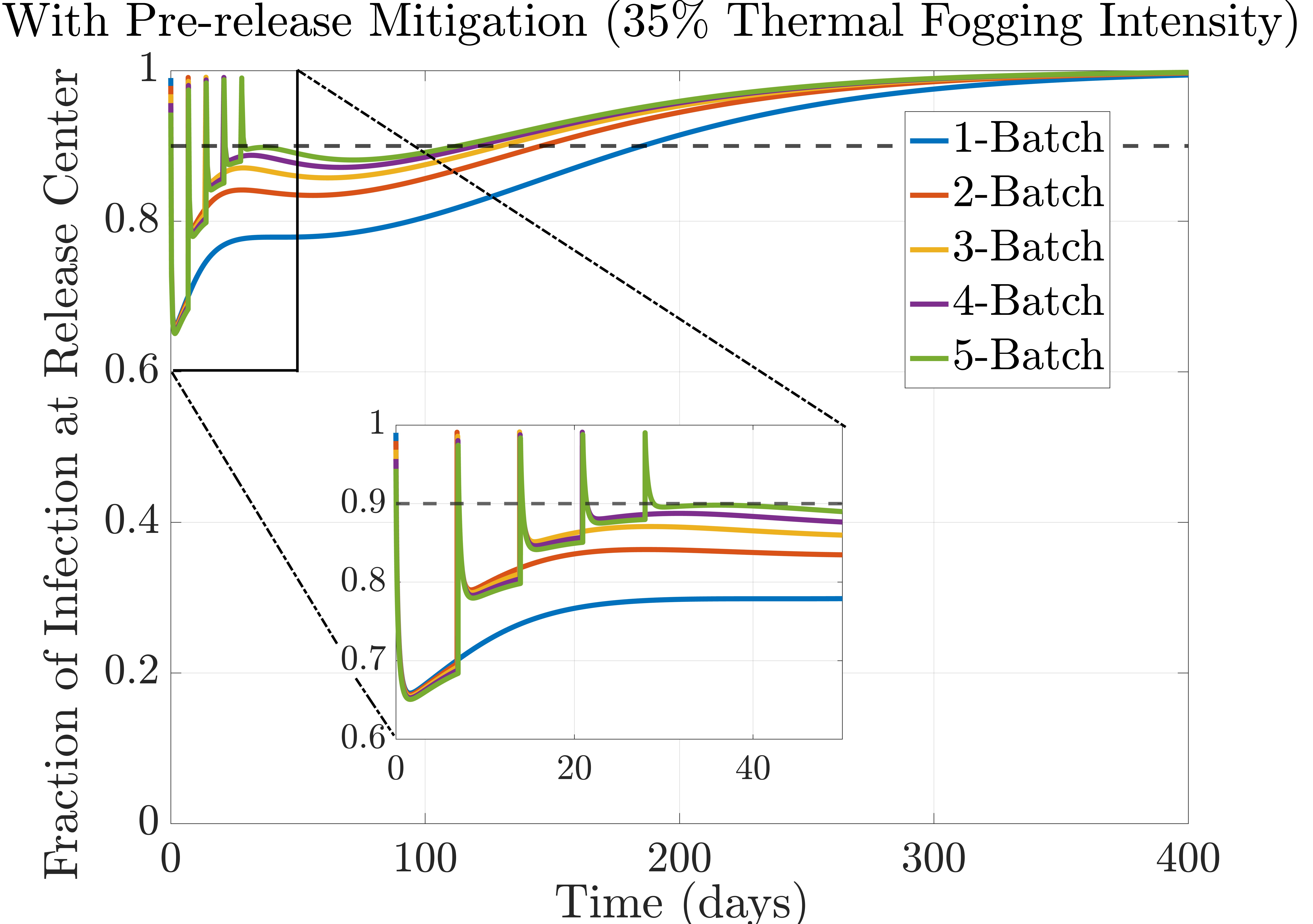}
   \vspace{.05in}
\end{minipage}
\caption{\small\textbf{Fraction of \textit{Wolbachia} infection over time (days)}. Performance comparison of releasing infected mosquitoes at the threshold level all at once or distributed in $2-5$ weekly batches. The impact of applying a pre-release mitigation strategy (thermal fogging $35\%$) was also evaluated. Domain size $L= 3000\;m\times 3000\;m$. Diffusion coefficient $D= 60\; m^2/$day. The standard deviation of Gaussian release of $\sigma_w=35$ meters.}
\label{fig: Mult_single_batch}
\end{figure}

Field trials often require periodic releases of batches of infected mosquitoes \cite{zheng2021one}. With the following simulations, we aim to study the impact of releasing a certain number of infected mosquitoes split into multiple batches released weekly over five consecutive weeks. All the releases had the same number of infected mosquitoes. Each batch of released mosquitoes was the total released size divided by the number of batches, and they were released at regular time intervals. The results of these simulations are summarized in Table \ref{tab: Mult-batch-release}, and the behavior of the corresponding fraction of infection curves can be visualized in Figure \ref{fig: Mult_single_batch}.

To evaluate the performance of these scenarios, we measured the number of days required for \textit{Wolbachia} to infect $90\%$ of the mosquito population. In the first scenario, shown in the left graph of Figure \ref{fig: Mult_single_batch}, no pre-release mitigation strategy was applied. We observe that releasing the infected mosquitoes in multiple batches ($2–5$ weekly batches) reduces the time to establishment to within a year, compared to releasing them all at once at the threshold level, where infection is not established within this time frame. 

On the other hand, in the right panel, we observe that if $35\%$ of the wild uninfected population is killed before releasing the mosquitoes in $2–5$ batches, the time to establish infection is reduced by approximately 23 days compared to the case with no pre-release intervention. A more significant reduction is seen in the single-release case, where thermal fogging speeds up the establishment by approximately 4 years. 

The results are summarized in Table \ref{tab: Mult-batch-release}. These findings highlight the importance of applying a pre-release mitigation strategy and incorporating a multiple-release approach in field trials to accelerate the establishment of \textit{Wolbachia} in the wild mosquito population.

\begin{table}[!ht]
\begin{adjustwidth}{-2.25in}{0in} 
\centering
\caption{
{\textbf{Simulation results when releasing \textit{Wolbachia}-infected mosquitoes in multiple batches}.Comparison of the number of days required to achieve $90\%$ infection when releasing \textit{Wolbachia}-infected mosquitoes at the threshold level ($1,072,640$ females and males) all at once or distributed in $2-5$ weekly batches over an area with a mosquito dispersal rate of $60\;m^2/$day. All five scenarios were evaluated with and without pre-release mitigation (Thermal Fogging $35\%$). The table shows the time to reach the threshold infection level for single-release and multi-batch mosquito releases. Pre-release mitigation significantly reduces the time to reach $90\%$ infection, particularly for the single-release strategy, where thermal fogging reduces the days to 185 compared to 5 years without mitigation. Multi-batch releases (2–5 batches) generally show a quicker progression towards reaching $90\%$ of infection, with the shortest times observed in the 5-batch release scenario, especially when combined with pre-release mitigation.\vspace{0.1in}}}
\begin{tabular}{|c|c|c|}
\hline
\textbf{No. Days to Achieve $90\%$ of Infection} & \textbf{No Pre-release Mitigation} & \textbf{With Pre-release Mitigation} \\ 
\textbf{} & \textbf{} & \textbf{(Thermal Fogging $35\%$)} \\ \hline
Single Release & 5 years & 185 days \\ \hline
2-Batch Release & 174 days & 146 days \\ \hline
3-Batch Release & 189 days & 145 days \\ \hline
4-Batch Release & 131 days & 120 days \\ \hline
5-Batch Release & 123 days & 113 days \\ \hline
\end{tabular}
\label{tab: Mult-batch-release}
\end{adjustwidth}
\end{table}

It is worth noticing that, as a result of modeling field releases with the spatial mosquito dispersion effect, the conclusions obtained from the previous simulations differ greatly from the multiple-release analysis performed in \cite{florez2023modeling} with an ODE model that assumes the fraction of infection among mosquitoes is homogeneous in space. This ODE model showed that without pre-release mitigation, releasing all the infection at once may not be as effective as splitting the release into multiple batches, which is consistent with our results. The benefit of leaving a gap between releases is related to the limited environmental resources available to sustain the uninfected population, not treated with any insecticide, as well as the introduction of additional infected mosquitoes, which overpopulate the region and increase the competition for resources when released in a single batch.

On the other hand, when there was pre-release mitigation, the ODE model results in \cite{florez2023modeling} suggest that releasing the infected mosquitoes all at once was more effective than splitting them into batches. However, in our simulations, we observe the opposite behavior, i.e., releasing the infection in batches could lead to a more effective strategy than releasing them all at once. Indeed, with our spatial model, where we assume a limited mosquito dispersal, represented by ranges with small diffusion coefficients $10-100$ $m^2/$day,  releasing \textit{Wolbachia} infected mosquitoes in batches allows for a more localized establishment of the infection within specific regions of the habitat. This is because mosquitoes interact primarily with nearby individuals, resulting in clusters of \textit{Wolbachia} infected mosquitoes in the regions where they were released. 

Now, suppose we add the assumption that a fraction of uninfected individuals are killed with thermal fogging before the release. This will decrease the infected cohort's competition for resources in these localized areas, creating a favorable environment for this population to breed and transmit \textit{Wolbachia} to their offspring.

\subsection{Sensitivity Analysis}\label{SA-regular}

The parameter values of the reduced model in Table \ref{tab: ParameterTable-Regular-Model} correspond to baseline estimates that carry uncertainty with respect to the biological measurements and vary across \textit{Wolbachia} strains, climatic conditions, mosquito species, etc. To account for this uncertainty level in our parameters, we use sensitivity analysis to quantify the relative significance of the model parameters of interest (POIs) toward the output quantities of interest (QOIs).

Based on the method followed in \cite{chitnis2008determining}, we define the normalized sensitivity index (SI) of a quantity of interest QOI, q(p), with respect to a parameter of interest POI, p, as

\begin{equation}
S_{p}^{q}=\frac{p}{q}\times \frac{\partial{q}}{\partial{p}}\Bigr|_{p=\hat{p}},
\end{equation}
evaluated at the baseline value $p=\hat{p}$. This dimensionless index is interpreted as the impact of percentage change: if the parameter of interest $p$ changes by $x\%$ around the baseline, then the quantity of interest $q$ changes by $S_{p}^{q}\times x\%.$ 

Given that our simulations employed the infection threshold ($p_{thres}$), computed in Section \ref{chatacterize-threshold}, as a foundational metric for designing effective \textit{Wolbachia} release strategies, we assess the sensitivity of this threshold to variations in the baseline values of biologically relevant parameters in the reduced $2$-PDE model used in our simulations. The results of this analysis are summarized in Table \ref{tab:sensitivity-analysis}.

\begin{table}[h!]
\small
\begin{adjustwidth}{-2.25in}{0in} 
\begin{center}
\caption{\textbf{Local sensitivity analysis of the threshold of infection $p_{thres}$, with respect changes in relevant model parameters}. Normalized sensitivity index $(SI)$ of key biological parameters in the reduced 2-PDE model with respect to the threshold of infection ($p_{thres}$), considered as the quantity of interest (QOI). The baseline values of the parameters are provided in parentheses. Positive and negative indices indicate parameters that increase or decrease the threshold, respectively. The sensitivity index was calculated by increasing each baseline quantity of the parameters of interest by $1\%$. The infection threshold is more sensitive to the per capita reproduction rate of infected and uninfected cohorts, followed by the release radius.}
\label{tab:sensitivity-analysis}
\begin{tabular}{|c|c|}
\hline
\multicolumn{2}{|c|}{\textbf{Threshold of Infection $p_{thres}$ (QOI)}} \\ \hline
\textbf{Parameter of Interest (POI)} & \textbf{Normalized Sensitivity Index} \\ \hline
Diffusion coefficient $D$ (Baseline $= 60\;m^2/$day) & $5.15$ \\ \hline
Standard deviation of Gaussian release (Baseline $=35\;m$) & $-8.56$ \\ \hline
Adjusted death rate for $F_u$ (Baseline $= 1/26.25\;\text{days}^{-1}$) & $-7.53$ \\ \hline
Adjusted death rate for $F_w$ (Baseline $=1/24.25\;\text{days}^{-1}$) & $7.53$ \\ \hline
Adjusted per capita reproduction rate for $F_u$ (Baseline $=7\;\text{days}^{-1}$) & $20.92$ \\ \hline
Adjusted per capita reproduction rate for $F_w$ (Baseline $=5.7\;\text{days}^{-1}$) & $-20.94$ \\ \hline
Pre-release Mitigation in Low Dispersal Regime (Baseline $= 35\%$ efficacy, $D=60\;m^2/$day) & $-0.44$ \\ \hline
Pre-release Mitigation in High Dispersal Regime (Baseline $= 35\%$ efficacy, $D=90\;m^2/$day) & $-4.04$ \\ \hline
\end{tabular}
\end{center}
\end{adjustwidth}
\end{table}

If the diffusion coefficient, representing mosquito dispersal, increments by $1\%$, the number of mosquitoes required for achieving a sustainable infection should increase by approximately $5\%$. Additionally, if the standard deviation of the Gaussian release, denoted by $\sigma_w$, increases by $1\%$, the infection threshold decreases by approximately $8\%$. This suggests that releasing mosquitoes over a wider area, compared to the baseline value, may require fewer \textit{Wolbachia}-infected mosquitoes to establish an infection.

An increase of $1\%$ in the death rate of uninfected mosquitoes leads to an approximate $7\%$ reduction in the infection threshold. At the same time, the same perturbation applied to the infected cohort results in a $7\%$ increase in the threshold. This latter outcome aligns with the fact that a shorter lifespan means infected mosquitoes die more quickly. To maintain or increase the number of infected mosquitoes in the environment, more infected mosquitoes would need to be released to compensate for the higher death rate, directly impacting the infection threshold.

The infection threshold is highly sensitive to the per capita reproduction rates of both \textit{Wolbachia}-infected and uninfected mosquito cohorts. A small increase in the reproduction rate of uninfected mosquitoes leads to a higher threshold of infection (by $20\%$). This suggests that when the uninfected mosquito population grows, it makes it more difficult for \textit{Wolbachia} to spread, potentially because the balance of infected and uninfected mosquitoes shifts, diluting the proportion of infected mosquitoes in the population and thereby raising the threshold for successful infection establishment. Conversely, the same increase in the reproduction rate of \textit{Wolbachia}-infected mosquitoes results in a lower infection threshold (by $20\%$). This indicates that increasing the reproductive capacity of infected mosquitoes facilitates the spread of \textit{Wolbachia}, necessitating fewer infected mosquitoes for successful establishment and thereby reducing the threshold of infection.

Regarding the percentage of thermal fogging efficacy as a parameter of interest, we observe two distinct behaviors in the threshold sensitivity depending on the mosquito dispersal regime. In regions with low mosquito dispersal ($D=60\;m^2/$day), the infection threshold is less sensitive to small increments in thermal fogging efficacy, resulting in a modest reduction of only $0.44\%$ in the threshold. This finding aligns with our simulation results, which suggest that in areas with limited mosquito dispersal, the use of insecticide may not be essential for successful \textit{Wolbachia} establishment due to the potential local concentration of \textit{Wolbachia}-infected mosquitoes after release. In contrast, for high mosquito dispersal regimes ($D=90\;m^2/$day), the threshold is more sensitive to thermal fogging efficacy, with a more substantial $4\%$ reduction in the threshold for the same small increment in efficacy. This is consistent with our simulation results, highlighting the need for pre-release intervention strategies to support infection establishment in regions with high mosquito dispersal rates. These findings emphasize the importance of considering the local ecological context, particularly the mosquito dispersal patterns when designing and implementing \textit{Wolbachia}-based control strategies. 

\section{Conclusions and Discussion}\label{Conclusions-Discussion}

This study confronts the ongoing challenge of mosquito-borne diseases, such as dengue, which remain a significant global public health concern, particularly in urban areas with high population densities. Traditional vector control methods often fail to produce sustained reductions in disease transmission, necessitating novel approaches. \textit{Wolbachia}, a bacterium that prevents mosquitoes from transmitting viruses, has emerged as a promising biological control method. While prior research has demonstrated the effectiveness of \textit{Wolbachia}-based strategies, implementing these approaches in field settings has been hindered by the lack of spatially dynamic models and guidance on practical release strategies. This work sought to bridge that gap, using mathematical modeling to improve the design and implementation of \textit{Wolbachia} releases.

The mathematical framework developed in this study, based on partial differential equations, simulates the spatial dynamics of mosquito populations and the spread of the Wolbachia infection. The model accounts for key ecological processes, such as mosquito dispersal, infection dynamics, and practical interventions like pre-release insecticide spraying.

Motivated by challenges encountered during large-scale releases where the \textit{Wolbachia} strain 
\textit{w}Mel failed to achieve full establishment—such as in deployment programs in Medell\'in, Colombia \cite{velez2023large}, and Rio de Janeiro, Brazil \cite{dos2022estimating}—the insights from this modeling effort offer a valuable roadmap for improving \textit{Wolbachia} deployment strategies.

Our analysis identifies strategies for optimizing \textit{Wolbachia}-infected mosquito releases under constraints on both release size and the efficacy level of insecticide used for pre-release interventions. We examine how these constraints interact with various landscape features influencing mosquito movement in urban environments. When release size is restricted, we find that in areas with limited mosquito dispersal $(10–50\;m^2/$day), such as backyards or gated communities, \textit{Wolbachia} mosquitoes remain concentrated locally after release and insecticide use may not be necessary as a pre-release intervention. In contrast, in regions with higher dispersal rates $(50–100\;m^2/$day), such as residential blocks or parks, the incorporation of pre-release interventions—specifically reducing at least $35\%$ of the wild mosquito population—is recommended to enhance \textit{Wolbachia} establishment within nine months. Furthermore, when the efficacy level of insecticide is limited to $35\%$, smaller release sizes, relative to the constrained release size, are sufficient to achieve $90\%$ \textit{Wolbachia} infection in low-dispersal areas $(10–45\;m^2/$day). However, larger releases are required in high-dispersal regions (greater than $45\;m^2/$day) to ensure successful \textit{Wolbachia} establishment.

Additionally, our spatial model predicts qualitatively different results than the previous ODE model. It demonstrates that distributing the \textit{Wolbachia} release size in smaller weekly batches ($2-5$ batches) is more effective than a single large release, even without any pre-release intervention. This highlights that, by incorporating mosquito dispersion, the model removes a simplifying assumption of the previous ODE models and enhances biological relevance, providing more accurate and reliable predictions for real-world applications.  

While this study offers valuable recommendations for improving dengue control, it is important to consider several model assumptions when interpreting the results. The current model is built on a simplified framework to capture the essential \textit{Wolbachia} dynamics, extending insights from previous analytical work while balancing mathematical tractability with modeling complexity. Specifically, the model approximates mosquito dispersal using diffusion processes, which assumes rare large-step movements. To account for more frequent large-step movements, more complex formulations of spatial terms, such as integro-differential equations or fractional derivatives \cite{campeau2022evolutionary,kwasnicki2017ten,podlubny1998fractional,hemme2010influence}, would be required. The model also assumes identical movement patterns for infected and uninfected mosquitoes. It does not account for environmental factors, such as wind, or differences in dispersal between sexes or infection status.

Additionally, the model focuses exclusively on adult female mosquitoes, omitting the aquatic stages of the mosquito life cycle. As a result, it does not account for pre-release interventions such as larviciding, which would require tracking aquatic-stage mosquitoes. Furthermore, the pre-release intervention considered here, thermal fogging, is assumed to selectively target uninfected mosquitoes without affecting \textit{Wolbachia}-infected individuals. 
While these assumptions simplify the intricate dynamics of real-world systems, they serve as reasonable approximations to guide field applications effectively. Our future work will enhance the model's biological realism by addressing these limitations and incorporating more realistic dispersal mechanisms, environmental factors, and multi-stage mosquito dynamics \cite{FlorezPineda2024,zhuolinqu2024multistage}.

The implications of this work extend beyond the immediate application to dengue. The modeling framework and strategies developed here broadly apply to controlling other vector-borne diseases, such as Zika, chikungunya, and malaria. By addressing the spatial complexities of mosquito populations and integrating practical interventions, this research provides a scalable and adaptable approach to improving public health in regions most affected by these diseases. The results highlight the transformative potential of combining biological control methods like \textit{Wolbachia} with spatially optimized, evidence-based strategies, offering a robust path forward for combating mosquito-borne diseases globally.

\section*{Acknowledgments}
ZQ was partially supported by the National Science Foundation award DMS-2316242. The funder had no role in study design, data collection and analysis, decision to publish, or manuscript preparation.

\appendix
\setcounter{figure}{0} 
\setcounter{equation}{0} 
\setcounter{table}{0}
\renewcommand\thefigure{\thesection.\arabic{figure}}   
\renewcommand\theequation{\thesection.\arabic{equation}}  
\renewcommand\thetable{\thesection.\arabic{table}}   

\section{Supporting Information}
\subsection{Cylindrically Symmetric System for 2-D Releases}\label{cyllindrically-symmetric-derivation}
To account for the radial expansion effect at which the \textit{Wolbachia} wave propagates, we simplify the model in \ref{eq: 2-PDE-system1-scaledKf}-\ref{eq: 2-PDE-system1-scaledKf} by assuming cylindrical symmetry of the solution. Therefore, the spatial diffusion is defined as $\Delta_r=\frac{\partial^2}{\partial r^2} + \frac{1}{r}\frac{\partial }{\partial r}$ . For this purpose, let us consider our state variables as $u=u(r,t)$ and $w=w(r,t)$ where $r = \sqrt{x^2+y^2}$ for a given point in space $(x,y)$ in a $2D$ domain. Then, the cylindrically symmetric system will be given by:
\begin{subequations}
\begin{align}
\frac{\partial u}{\partial t} &= b_f\phi_u^{r}\frac{u}{u + \frac{\mu_{fw}^{r}}{\mu_{fu}^{r}}w}(1-u-w)u-\mu_{fu}^{r}u +D\Bigg(\frac{\partial u^2}{\partial r^2}+ \frac{1}{r}\frac{\partial u}{\partial r}\Bigg),\label{eq: 2-PDE-system1-polar-ap}\\
\frac{\partial w}{\partial t} &= b_f\phi_w^{r}(1-u-w)w - \mu_{fw}^{r}w + D\Bigg(\frac{\partial w^2}{\partial r^2}+ \frac{1}{r}\frac{\partial w}{\partial r}\Bigg). \label{eq: 2-PDE-system2-polar-ap}
\end{align}
\end{subequations}
To ensure as smooth transition across the singularity at $r=0$, we impose Neumann boundary conditions, i.e.,
\begin{subequations}
\begin{align}
\frac{\partial u}{\partial r}\big|_{r=0} = 0,\;\; \frac{\partial u}{\partial r}\big|_{r=L}=0,\label{eq: Neumann-bdy-cond1}\\
\frac{\partial w}{\partial r}\big|_{r=0} = 0,\;\; \frac{\partial w}{\partial r}\big|_{r=L}=0, \label{eq: Neumann-bdy-cond2}
\end{align}
\end{subequations}
where $L$ represents the radius of our circular domain. Note that this transformation of coordinates simplifies the 2-dimensional system in space into a 1-dimensional system for simulating a circular-shaped release protocol.
\subsection{Numerical Approximation of the 2-PDE System}\label{numerical-method}
The solution of the system of partial differential equations in \label{eq: 2-PDE-system1-polar}-\label{eq: 2-PDE-system2-polar} is approximated using a centered second-order finite difference method. Given a circular domain of radius $L$, we will denote $u_j^{n}\approx u(r_j,t_n)$ and $v_j^{n}\approx (r_j,t_n)$ as the approximate solutions of the state variables at a radial grid point $r_j\in[0,L]$ and time $t_n\in[0,T]$, where $T$ denotes the final time of the simulation. Assuming $N$ is the number of subintervals at which we divide the domain $[0,L]$, the spatial grid point is defined as $r_j=(j-1/2)h$, where $h=L/N$, for $j=1,\dots,N$. Thus, the system can be discretized as follows:
\begin{subequations}
\begin{align}
u_j^{n+1} &\approx u_j^n + \Delta t \Big(b_f\phi_u\Bigg(\frac{u_j^n}{u_j^n+\big(\frac{\mu_{fw}}{\mu_{fu}}\big)v_j^n}\Bigg)(1-u_j^n-v_j^n)u_j^n-\mu_{fu}u_j^n+D\Delta_r u_j^n  \Big),\label{eq: num_method_next1}  \\
v_j^{n+1} &\approx v_j^n + \Delta t \Big(b_f\phi_w(1-u_j^n-v_j^n)v_j^n-\mu_{fw}v_j^n+D \Delta_r v_j^n \Big). \label{eq: num_method_next2} 
\end{align}
\end{subequations}

\begin{figure}[hbtp]
	\centering
	\includegraphics[width=0.5\textwidth]{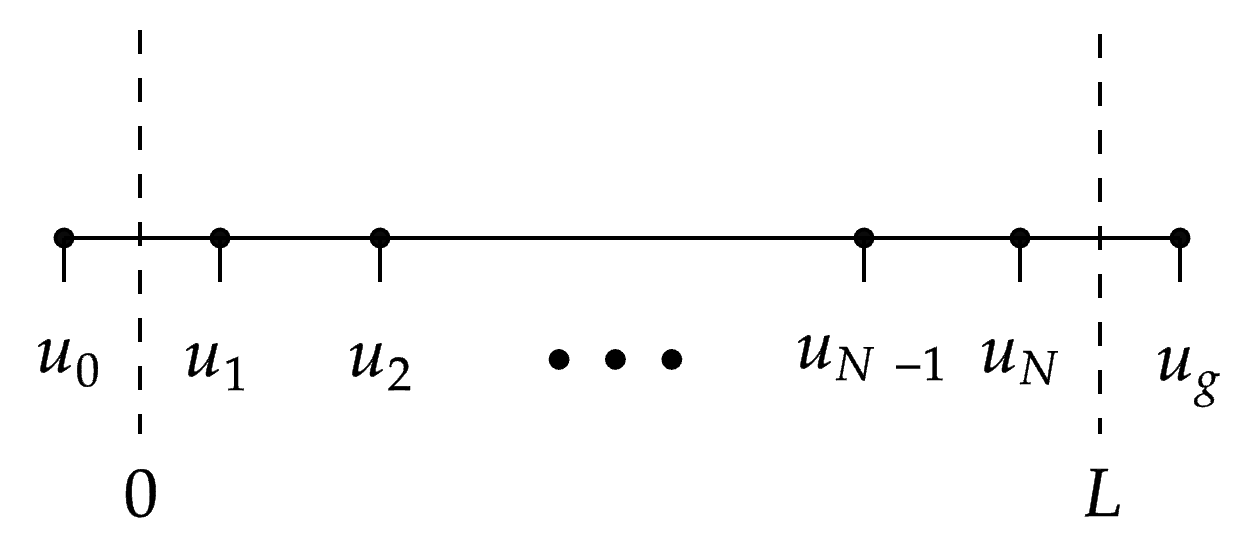}
	\caption{Discretization diagram for computing the solution using centered differences, where $u_g$ corresponds to a ghost point. We adjusted the grid points so that there is no grid point at $r=0$, avoiding the singularity.}
	\label{fig: Discretization-diagram}
\end{figure}

Here the time derivative is discretized using a Forward Euler Method, while the spatial derivative, which is present in the radial Laplacian term, is computed using centered differences:
\begin{subequations}
\begin{align}
\Delta_r u_j^n &\approx \frac{u_{j+1}^{n}-2u_{j}^{n}+u_{j-1}^{n}}{h^2} + \frac{1}{r_j}\Bigg(\frac{u_{j+1}^{n}-u_{j-1}^n}{2h}\Bigg) + O(h^2), \label{eq: lap_centered_diff_u}\\
\Delta_r v_j^{n} &\approx \frac{v_{j+1}^{n}-2v_{j}^{n}+v_{j-1}^{n}}{h^2} + \frac{1}{r_j}\Bigg(\frac{v_{j+1}^{n}-v_{j-1}^{n}}{2h}\Bigg) + O(h^2). \label{eq: lap_centered_diff_v}
\end{align}
\end{subequations}
The Neumann boundary condition at $r=0$ can be enforced by assuming symmetry of the functions $u$ and $w$ about the origin. More precisely,  the equations in \ref{eq: Neumann-bdy-cond1}-\ref{eq: Neumann-bdy-cond2} can be discretized as:
\begin{subequations}
\begin{align}
\frac{u_{1}^{n}-u_{0}^{n}}{2h} =0,\;\;\frac{u_{N+1}^{n}-u_{N}^{n}}{2h} =0,\\
\frac{v_{1}^{n}-v_{0}^{n}}{2h} =0,\;\;\frac{v_{N+1}^{n}-v_{N}^{n}}{2h}=0,
\end{align}
\end{subequations}
which imply $u_{0}^{n}=u_{1}^{n}$, $v_{0}^{n}=v_{1}^{n}$, $u_{N+1}^{n}=u_{N}^{n}$, and $v_{N+1}^{n}=v_{N}^{n}$.
In Figure \ref{fig: Discretization-diagram}, we present a graphical description of the spatial discretization implemented in our solver.
Note that with this discretization, the Laplacian at the boundary points $r=0$ and $r=L$ will be computed based on the following cases. For $j=1$:
\begin{subequations}
\begin{align}
\Delta u_{1}^n &= \frac{u_2^n-2u_1^n+u_0^n}{h^2} + \frac{u_2^n-u_0^n}{2hr_1}\\
&= \frac{u_2^n-u_1^n}{h^2}+ \frac{u_2^n-u_1^n}{2hr_1}.
\end{align}
\end{subequations}
Similarly, we can compute $\Delta v_1^n$. Now, for $j=N$, we have
\begin{subequations}
\begin{align}
\Delta u_{N}^n &= \frac{u_N^n-2u_N^n+u_{N-1}^n}{h^2} + \frac{u_N^n-u_{N-1}^n}{2hr_N},\\
\Delta v_{N}^n &= \frac{v_N^n-2v_N^n+v_{N-1}^n}{h^2} + \frac{v_N^n-v_{N-1}^n}{2hr_N}.
\end{align}
\end{subequations}
Regarding the initial conditions of the system, we define them as follows:
\begin{subequations}
\begin{align}
u(r,0) &= c_0,\\
w(r,0) &= ae^{-r^2/b^2}.
\end{align}
\end{subequations}
Here, $u(r,0)$ will be a constant function with the value $c_0$, which is determined by the carrying capacity of the mosquito population. Similarly, $w(r,0)$ will be an inverse squared exponential function with real constants $a$ and $b \neq 0$. The parameter $a$ represents the height of the peak, while $R$ controls the radius of the release shape. These two parameters also depend on the carrying capacity and determine the number of mosquitoes to be released in the field. 
\subsection{Convergence Test of Numerical Approximation of Threshold of Infection}\label{convergence-test-threshold}
\begin{figure}[h]
  \centering
  \includegraphics[width=.7\linewidth]{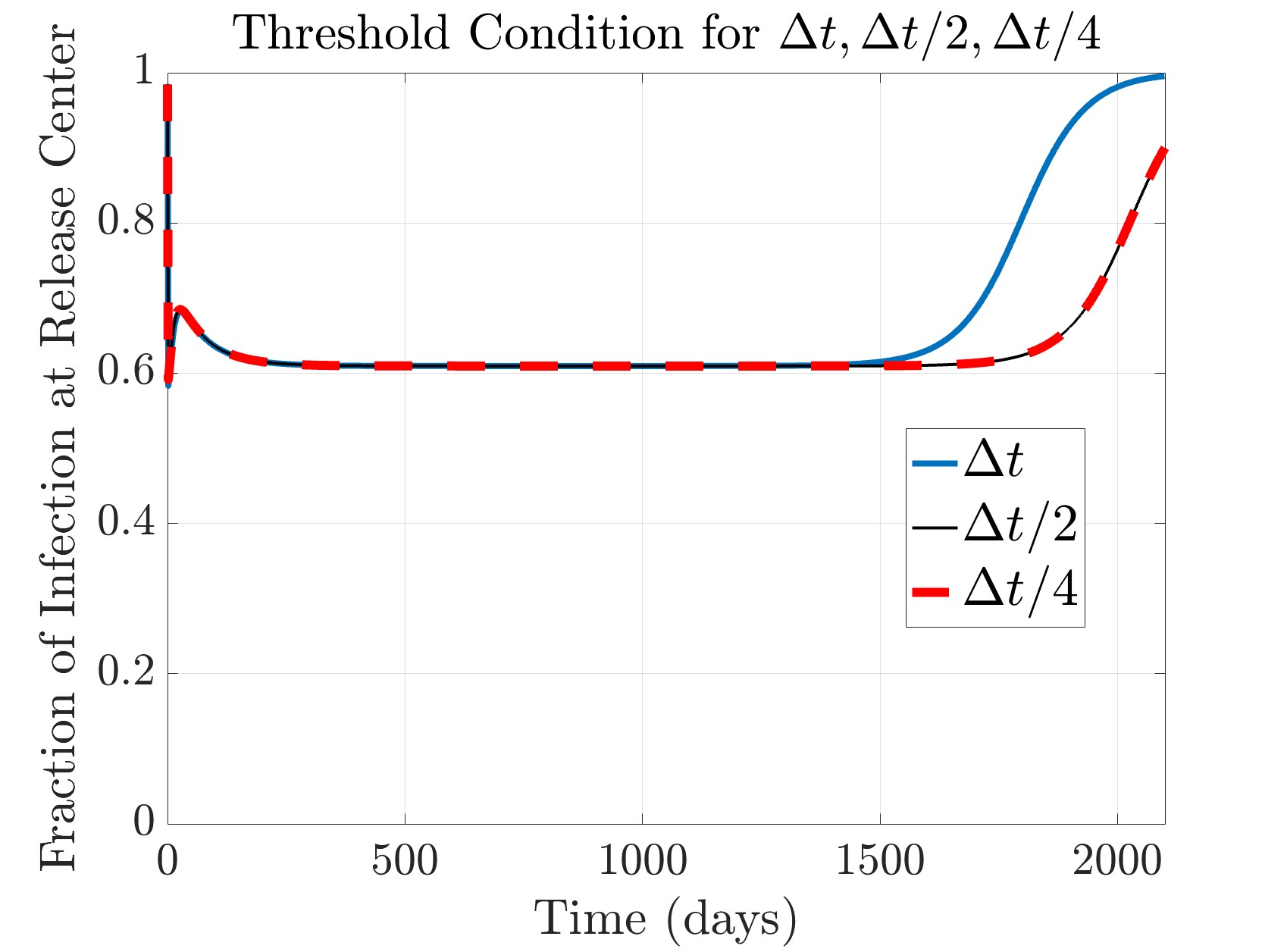}
\caption{\small Fraction of infection curves generated by releasing \textit{Wolbachia}-infected mosquitoes at the threshold conditions determined by three PDE-solver time steps, $\Delta t$, $\Delta t/2$ and $\Delta t/4$, where the baseline time step value is $\Delta t= 1.78\times10^{-2}$. As the time-step decreases, the infection curves converge to a single shape whose initial condition corresponds to the numerical approximation of the threshold condition.}
\label{fig: Threshold-time-step}
\end{figure}

The root-finding algorithm presented in section \ref{chatacterize-threshold} to characterize the threshold condition is sensitive to the time-step of the PDE solver. In Table \ref{tab:threshold_vs_timestep}, we have computed $p_{thres}$ for three different time steps that differ consecutively by a factor of $1/2$. We observe that the threshold condition among consecutive time steps reduces significantly and the infection curves will start to approach a single shape (see Figure \ref{fig: Threshold-time-step}).

\begin{table}[!ht]
\begin{adjustwidth}{-2.25in}{0in}
\centering
\caption{
{\textbf{Convergence test of the numerical approximation of the threshold of infection} Threshold values for three different PDE solver time steps. These values are expressed as the total number of infected mosquitoes released and their corresponding fraction of infection with respect to the female carrying capacity $K_f$. As the time step decreases by $50\%$, the difference between consecutive threshold values decreases.\vspace{0.1in}}}
\begin{tabular}{|c|c|c|c|}
\hline
\textbf{Scaling Factor} & \textbf{Time Step} & \textbf{Total No. of Mosquitoes Released} & \textbf{Initial Fraction of Infection $p_{thres}$} \\ \hline
$\Delta t$    & 0.017 & 1,072,640 & 0.357 \\ \hline
$\Delta t/2$  & 0.008 & 1,047,040 & 0.349 \\ \hline
$\Delta t/4$  & 0.004 & 1,032,140 & 0.344 \\ \hline
\end{tabular}
\label{tab:threshold_vs_timestep}
\end{adjustwidth}
\end{table}

From these computations, it is possible to estimate the limiting value of the threshold condition as $\Delta t \rightarrow 0$. Let $p^*$ denote this limiting value, then we can express the threshold conditions of Table \ref{tab:threshold_vs_timestep} in terms of $p^*$ as follows:
\begin{subequations}
\begin{align}
p_{thres}\Big(\Delta t\Big) &= p^* + C\Big(\Delta t\Big)^m, \label{eq: threshold_deltat}\\
p_{thres}\Big(\frac{\Delta t}{2}\Big) &= p^* + C\Big(\frac{\Delta t}{2}\Big)^m, \label{eq: threshold_deltat2}\\
p_{thres}\Big(\frac{\Delta t}{4}\Big) &= p^* + C\Big(\frac{\Delta t}{4}\Big)^m, \label{eq: threshold_deltat4}
\end{align}
\end{subequations}
for some power $m>0$ and some constant $C>0$. Now, if we compute the ratio difference among consecutive threshold values, we get:
\begin{subequations}
\begin{align}
p_{thres}\Big( \Delta t \Big)-p_{thres}\Big(\frac{\Delta t}{2}\Big)&= C(\Delta t)^m(2^m-1)2^{-m},\\
p_{thres}\Big(\frac{\Delta t}{2}\Big)-p_{thres}\Big(\frac{\Delta t}{4}\Big)&= C(\Delta t)^m2^{-m}(2^m-1)2^{-m},\\
\frac{p_{thres}\Big(\Delta t\Big)-p_{thres}\Big(\frac{\Delta t}{2}\Big)}{p_{thres}\Big(\frac{\Delta t}{2}\Big)-p_{thres}\Big(\frac{\Delta t}{4}\Big)} &= \frac{1}{2^{-m}} = 2^m.
\end{align}
\end{subequations}
From the previous equations, we can deduce the values for $m$ and $C$:
\begin{subequations}
\begin{align}
m &= log_{2}\Bigg( \frac{p_{thres}(\Delta t)-p_{thres}(\Delta t/2)}{p_{thres}(\Delta t/2)-p_{thres}(\Delta t/4)}\Bigg) \approx 0.6781,\\
C &= \frac{p_{thres}(\Delta t)-p_{thres}(\Delta t/2)}{(\Delta t)^m(2^m-1)2^{-m}} \approx 0.3277
\end{align}
\end{subequations}
Now, if we replace the values of $m$ and $C$ in \ref{eq: threshold_deltat}, we obtain an estimation of $p^{*}$:
\begin{align}
p^*=p_{thres}(\Delta t) - C(\Delta t)^m &\approx 0.3357\;\; \text{(fraction of infection)}\\ 
& \approx 1,007,100\;\;\text{(total number of mosquitoes)}
\end{align}


\begin{thebibliography}{10}

\bibitem{franklinos2019effect}
Franklinos LH, Jones KE, Redding DW, Abubakar I.
\newblock The effect of global change on mosquito-borne disease.
\newblock The Lancet infectious diseases. 2019;19(9):e302--e312.

\bibitem{manore2014comparing}
Manore CA, Hickmann KS, Xu S, Wearing HJ, Hyman JM.
\newblock Comparing dengue and chikungunya emergence and endemic transmission in A. aegypti and A. albopictus.
\newblock Journal of theoretical biology. 2014;356:174--191.

\bibitem{qu2018modeling}
Qu Z, Xue L, Hyman JM.
\newblock Modeling the transmission of Wolbachia in mosquitoes for controlling mosquito-borne diseases.
\newblock SIAM Journal on Applied Mathematics. 2018;78(2):826--852.

\bibitem{qu2022modeling}
Qu Z, Wu T, Hyman JM.
\newblock Modeling spatial waves of Wolbachia invasion for controlling mosquito-borne diseases.
\newblock SIAM Journal on Applied Mathematics. 2022;82(6):1903--1929.

\bibitem{barton2011spatial}
Barton NH, Turelli M.
\newblock Spatial waves of advance with bistable dynamics: cytoplasmic and genetic analogues of Allee effects.
\newblock The American Naturalist. 2011;178(3):E48--E75.

\bibitem{lewis1993waves}
Lewis M, Van Den~Driessche P.
\newblock Waves of extinction from sterile insect release.
\newblock Mathematical Biosciences. 1993;116(2):221--247.

\bibitem{qu2019generating}
Qu Z, Hyman JM.
\newblock Generating a hierarchy of reduced models for a system of differential equations modeling the spread of Wolbachia in mosquitoes.
\newblock SIAM Journal on Applied Mathematics. 2019;79(5):1675--1699.

\bibitem{Qu_Wu_Hyman_2022}
Qu Z, Wu T, Hyman JM.
\newblock Modeling Spatial Waves of Wolbachia Invasion for Controlling Mosquito-Borne Diseases.
\newblock SIAM Journal on Applied Mathematics. 2022;82(6):1903–1929.
\newblock doi:{10.1137/21M1440384}.

\bibitem{turner2023economic}
Turner HC, Quyen DL, Dias R, Huong PT, Simmons CP, Anders KL.
\newblock An economic evaluation of Wolbachia deployments for dengue control in Vietnam.
\newblock PLOS Neglected Tropical Diseases. 2023;17(5):e0011356.

\bibitem{utarini2021efficacy}
Utarini A, Indriani C, Ahmad RA, Tantowijoyo W, Arguni E, Ansari MR, et~al.
\newblock Efficacy of Wolbachia-infected mosquito deployments for the control of dengue.
\newblock New England Journal of Medicine. 2021;384(23):2177--2186.

\bibitem{o2018scaled}
O'Neill SL, Ryan PA, Turley AP, Wilson G, Retzki K, Iturbe-Ormaetxe I, et~al.
\newblock Scaled deployment of Wolbachia to protect the community from dengue and other Aedes transmitted arboviruses.
\newblock Gates open research. 2018;2.

\bibitem{indriani2020reduced}
Indriani C, Tantowijoyo W, Ranc{\`e}s E, Andari B, Prabowo E, Yusdi D, et~al.
\newblock Reduced dengue incidence following deployments of Wolbachia-infected Aedes aegypti in Yogyakarta, Indonesia: a quasi-experimental trial using controlled interrupted time series analysis.
\newblock Gates open research. 2020;4.

\bibitem{pinto2021effectiveness}
Pinto SB, Riback TI, Sylvestre G, Costa G, Peixoto J, Dias FB, et~al.
\newblock Effectiveness of Wolbachia-infected mosquito deployments in reducing the incidence of dengue and other Aedes-borne diseases in Niter{\'o}i, Brazil: A quasi-experimental study.
\newblock PLoS neglected tropical diseases. 2021;15(7):e0009556.

\bibitem{dos2022estimating}
Dos~Santos GR, Durovni B, Saraceni V, Riback TIS, Pinto SB, Anders KL, et~al.
\newblock Estimating the effect of the wMel release programme on the incidence of dengue and chikungunya in Rio de Janeiro, Brazil: a spatiotemporal modelling study.
\newblock The Lancet Infectious Diseases. 2022;22(11):1587--1595.

\bibitem{velez2023large}
Velez ID, Uribe A, Barajas J, Uribe S, {\'A}ngel S, Suaza-Vasco JD, et~al.
\newblock Large-scale releases and establishment of w Mel Wolbachia in Aedes aegypti mosquitoes throughout the Cities of Bello, Medell{\'\i}n and Itag{\"u}{\'\i}, Colombia.
\newblock PLOS Neglected Tropical Diseases. 2023;17(11):e0011642.

\bibitem{calle2024evaluation}
Calle-Tob{\'o}n A, Rojo-Ospina R, Zuluaga S, Giraldo-Mu{\~n}oz JF, Cadavid JM.
\newblock Evaluation of Wolbachia infection in Aedes aegypti suggests low prevalence and highly heterogeneous distribution in Medell{\'\i}n, Colombia.
\newblock Acta Tropica. 2024;260:107423.

\bibitem{takahashi2005mathematical}
Takahashi LT, Maidana NA, Ferreira WC, Pulino P, Yang HM.
\newblock Mathematical models for the Aedes aegypti dispersal dynamics: traveling waves by wing and wind.
\newblock Bulletin of Mathematical Biology. 2005;67:509--528.

\bibitem{walker2011w}
Walker T, Johnson P, Moreira L, Iturbe-Ormaetxe I, Frentiu F, McMeniman C, et~al.
\newblock The w Mel Wolbachia strain blocks dengue and invades caged Aedes aegypti populations.
\newblock Nature. 2011;476(7361):450--453.

\bibitem{laven1967eradication}
Laven H.
\newblock Eradication of Culex pipiens fatigans through cytoplasmic incompatibility.
\newblock Nature. 1967;216(5113):383--384.

\bibitem{rejmankova2013ecology}
Rejm{\'a}nkov{\'a} E, Grieco J, Achee N, Roberts DR.
\newblock Ecology of larval habitats.
\newblock Anopheles mosquitoes-New insights into malaria vectors. 2013; p. 397--446.

\bibitem{durovni2020impact}
Durovni B, Saraceni V, Eppinghaus A, Riback TI, Moreira LA, Jewell NP, et~al.
\newblock The impact of large-scale deployment of Wolbachia mosquitoes on dengue and other Aedes-borne diseases in Rio de Janeiro and Niter{\'o}i, Brazil: study protocol for a controlled interrupted time series analysis using routine disease surveillance data.
\newblock F1000Research. 2020;8:1328.

\bibitem{fife2013mathematical}
Fife PC.
\newblock Mathematical aspects of reacting and diffusing systems. vol.~28.
\newblock Springer Science \& Business Media; 2013.

\bibitem{russell2005mark}
Russell RC, Webb C, Williams C, Ritchie S.
\newblock Mark--release--recapture study to measure dispersal of the mosquito Aedes aegypti in Cairns, Queensland, Australia.
\newblock Medical and veterinary entomology. 2005;19(4):451--457.

\bibitem{winskill2015dispersal}
Winskill P, Carvalho DO, Capurro ML, Alphey L, Donnelly CA, McKemey AR.
\newblock Dispersal of engineered male Aedes aegypti mosquitoes.
\newblock PLoS neglected tropical diseases. 2015;9(11):e0004156.

\bibitem{filipovic2020using}
Filipovi{\'c} I, Hapuarachchi HC, Tien WP, Razak MABA, Lee C, Tan CH, et~al.
\newblock Using spatial genetics to quantify mosquito dispersal for control programs.
\newblock BMC biology. 2020;18:1--15.

\bibitem{sikorski1982bisection}
Sikorski K.
\newblock Bisection is optimal.
\newblock Numerische Mathematik. 1982;40:111--117.

\bibitem{hu2021mosquito}
Hu L, Yang C, Hui Y, Yu J.
\newblock Mosquito control based on pesticides and endosymbiotic bacterium Wolbachia.
\newblock Bulletin of Mathematical Biology. 2021;83(5):58.

\bibitem{hoffmann2013facilitating}
Hoffmann AA, Turelli M.
\newblock Facilitating Wolbachia introductions into mosquito populations through insecticide-resistance selection.
\newblock Proceedings of the Royal Society B: Biological Sciences. 2013;280(1760):20130371.

\bibitem{zheng2021one}
Zheng B, Li J, Yu J.
\newblock One discrete dynamical model on the Wolbachia infection frequency in mosquito populations.
\newblock Science China Mathematics. 2021; p. 1--16.

\bibitem{hancock2011strategies}
Hancock PA, Sinkins SP, Godfray HCJ.
\newblock Strategies for introducing Wolbachia to reduce transmission of mosquito-borne diseases.
\newblock PLoS neglected tropical diseases. 2011;5(4):e1024.

\bibitem{florez2023modeling}
Florez D, Young AJ, Bernab{\'e} KJ, Hyman JM, Qu Z.
\newblock Modeling sustained transmission of Wolbachia among Anopheles mosquitoes: Implications for malaria control in Haiti.
\newblock Tropical Medicine and Infectious Disease. 2023;8(3):162.

\bibitem{goindin2015parity}
Goindin D, Delannay C, Ramdini C, Gustave J, Fouque F.
\newblock Parity and longevity of Aedes aegypti according to temperatures in controlled conditions and consequences on dengue transmission risks.
\newblock PloS one. 2015;10(8):e0135489.

\bibitem{hoffmann2011successful}
Hoffmann AA, Montgomery B, Popovici J, Iturbe-Ormaetxe I, Johnson P, Muzzi F, et~al.
\newblock Successful establishment of Wolbachia in Aedes populations to suppress dengue transmission.
\newblock Nature. 2011;476(7361):454--457.

\bibitem{longdistancesCDC}
for Emerging NC, (NCEZID) ZID. Life Cycle of Aedes Mosquitoes;.
\newblock \url{https://www.cdc.gov/mosquitoes/about/life-cycle-of-aedes-mosquitoes.html#print}.

\bibitem{reiter1995dispersal}
Reiter P, Amador MA, Anderson RA, Clark GG.
\newblock Short Report: Dispersal of Aedes aegypti in an urban area after blood feeding as demonstrated by rubidium-marked eggs.
\newblock The American journal of tropical medicine and hygiene. 1995;52(2):177--179.

\bibitem{schmidt2017local}
Schmidt TL, Barton NH, Ra{\v{s}}i{\'c} G, Turley AP, Montgomery BL, Iturbe-Ormaetxe I, et~al.
\newblock Local introduction and heterogeneous spatial spread of dengue-suppressing Wolbachia through an urban population of Aedes aegypti.
\newblock PLoS biology. 2017;15(5):e2001894.

\bibitem{de2017does}
de~Oliveira S, Villela DAM, Dias FBS, Moreira LA, Maciel~de Freitas R.
\newblock How does competition among wild type mosquitoes influence the performance of Aedes aegypti and dissemination of Wolbachia pipientis?
\newblock PLoS Neglected Tropical Diseases. 2017;11(10):e0005947.

\bibitem{chitnis2008determining}
Chitnis N, Hyman JM, Cushing JM.
\newblock Determining important parameters in the spread of malaria through the sensitivity analysis of a mathematical model.
\newblock Bulletin of mathematical biology. 2008;70:1272--1296.

\bibitem{campeau2022evolutionary}
Campeau W, Simons AM, Stevens B.
\newblock The evolutionary maintenance of L{\'e}vy flight foraging.
\newblock PLoS computational biology. 2022;18(1):e1009490.

\bibitem{kwasnicki2017ten}
Kwa{\'s}nicki M.
\newblock Ten equivalent definitions of the fractional Laplace operator.
\newblock Fractional Calculus and Applied Analysis. 2017;20(1):7--51.

\bibitem{podlubny1998fractional}
Podlubny I.
\newblock Fractional differential equations: an introduction to fractional derivatives, fractional differential equations, to methods of their solution and some of their applications.
\newblock elsevier; 1998.

\bibitem{hemme2010influence}
Hemme RR, Thomas CL, Chadee DD, Severson DW.
\newblock Influence of urban landscapes on population dynamics in a short-distance migrant mosquito: evidence for the dengue vector Aedes aegypti.
\newblock PLoS neglected tropical diseases. 2010;4(3):e634.

\bibitem{FlorezPineda2024}
R C, D F.
\newblock Mathematical Models for Transmission and Control of Mosquito-borne Diseases.
\newblock School of Science \& Engineering Mathematics Degree granting institution. July, 2024;.

\bibitem{zhuolinqu2024multistage}
{Zhuolin Qu}, {Tong Wu}.
\newblock Multistage {{Spatial Model}} for {{Informing Release}} of {{Wolbachia-infected Mosquitoes}} as {{Disease Control}}.
\newblock Submitted. 2024;.

\end{thebibliography}
\end{document}